\documentstyle[12pt,graphicx,epsf]{article}
\begin{document}

\newcommand{\bev}{\begin{verbatim}}
\newcommand{\beq}{\begin{equation}}
\newcommand{\beqa}{\begin{eqnarray}}
\newcommand{\beqn}{\begin{eqnarray}}
\newcommand{\eeqn}{\end{eqnarray}}
\newcommand{\eeqa}{\end{eqnarray}}
\newcommand{\eeq}{\end{equation}}
\newcommand{\Eev}{\end{verbatim}}
\newcommand{\bec}{\begin{center}}
\newcommand{\eec}{\end{center}}
\def\ie{{\it i.e.}}
\def\eg{{\it e.g.}}
\def\half{{\textstyle{1\over 2}}}
\def\nicefrac#1#2{\hbox{${#1\over #2}$}}
\def\third{{\textstyle {1\over3}}}
\def\quarter{{\textstyle {1\over4}}}
\def\m{{\tt -}}
\def\mass{M_{l^+ l^-}}
\def\p{{\tt +}}

\def\slash#1{#1\hskip-6pt/\hskip6pt}
\def\slk{\slash{k}}
\def\GeV{\,{\rm GeV}}
\def\TeV{\,{\rm TeV}}
\def\y{\,{\rm y}}

\def\l{\langle}
\def\r{\rangle}

\begin{titlepage}
\samepage{
\setcounter{page}{1}
\rightline{OUTP-01-05P}
\rightline{UNILE-CBR-2001-2}

\vspace{1.5cm}
\begin{center}
 {\Large \bf   Supersymmetric Scaling Violations (II).\\
  The general supersymmetric Evolution \\  }
\vspace{1cm}
 {\large Claudio Corian\`{o}\footnote{ E-mail address: Claudio.Coriano@le.infn.it}\\   

\vspace{1cm}
{\it Theoretical Physics Department\\
University of Oxford, Oxford, OX1 3NP, United Kingdom}
\vspace{.5cm}
\centerline{and}
\vspace{.5cm}
\footnote{ Permanent Address}
{\it Dipartimento di Fisica, Universita' di Lecce\\
I.N.F.N. Sezione di Lecce \\
Via Arnesano, 73100 Lecce, Italy}}
\end{center}
\vspace{.01in}

\begin{abstract}
Using a recursive algorithm to solve the renormalization group equations
of $N=1$ QCD (DGLAP), we describe the most general supersymmetric evolution 
of the parton distributions. The analysis involves 
the regular DGLAP evolution, a partial supersymmetric intermediate evolution
and a final supersymmetric evolution combined at various matching scales.
We use a model in which supersymmetric distributions are 
radiatively generated at each susy threshold 
due to the mixing of the QCD anomalous dimensions with the $N=1$ sector. 
Various types of matching conditions are considered, 
reflecting partially broken and exact supersymmetric 
scenarios with a wide range of susy-breaking mass parameters. 
Numerical results for all the distributions are presented. 
\end{abstract}
\smallskip}
\end{titlepage}

\section{Introduction}
The search for viable supersymmetric extensions of the Standard Model has acquired a new momentum with the coming into operation of the Large Hadron Collider at Cern, scheduled for the near future. 
A great deal of work is currently undertaken in trying to 
map the various scenarios that supersymmetric models predicts in this unexplored energy range, 
by providing estimates for the various channels 
which may become available at this new scale.
Being the symmetry broken at such energy, 
the several parameters describing the breaking make its description 
a complex task. 
In this respect, there are different avenues that these studies 
can take. For instance, one possibility is to investigate supersymmetric 
corrections 
affecting the initial state, prior to the hard scattering phase;  
or to analize the opening of heavy supersymmetric channels in the 
intermediate/final state of the collisions; or, finally, 
to investigate a combination of both. 
These avenues are all - although not equally - justified by current 
experimental lower bounds on the mass of the supersymmetric partners.  
Supersymmetric scaling violations, which are those considered in this work,
 have to do with the first type of searches. They have also to do with a
regime of the theory in which the symmetry is - to some extent - restored, and all the 
channels become, effectively,  massless. We will elaborate on them in some detail in the following sections.  

We have presented in a previous paper results for the
evolution of regular and supersymmetric parton distributions within a
scenario characterized by broken supersymmetry and a decoupled squark. 
In this work we intend to examine this subject from a different
perspective. We analize an exact supersymmetric phase of the theory and
study the structure of the distribution of quarks and gluons and of
their supersymmetric partners, gluinos and squarks, under these
conditions. This scenario is, obviously, less realistic than a 
scenario characterized by a broken susy, but not less interesting
neverthless. In
particular, we will present some comparisons between the regular QCD
scenario, various types of susy breaking scenarios and the exact
supersymmetric scenario. 

The aim of this analysis is to quantify these
effects with a reasonable accuracy, in such a way that the results can be used
as a guideline for experimental searches in the future. We build on
previous old work of Kounnas and Ross \cite{KR} who calculated the
leading order anomalous dimensions for the evolution of supersymmetric
QCD (Susy DGLAP) and analized the first 2 moments of the distributions. 
Our phenomenological analysis is updated to current estimates of the
parameters of the parton distributions. It includes all the
moments, since it is an exact 
numerical solution of the evolution performed by standard iteration of
the convolution integrals, recast in the form of a recursion relation.  
In the section that follows, we will use the notation AP to denote the standard DGLAP 
evolution. The acronym SDGLAP or, briefly, SAP, will denote the partial supersymmetric 
DGLAP evolution with coupled gluinos and decoupled squarks. Finally, the acronym 
ESDGLAP or ESAP will denote the exact supersymmetric DGLAP for $N=1$ QCD 
with coupled gluinos {\em and} squarks.  

 \section{The Evolution Equations of SQCD with Exact Supersymmetry (ESAP)}
We refer to previous work of us for a detailed description of the
algorithm that we are going to use in this analysis \cite{CC1}. We introduce some
definitions just in order to make our discussion self contained.  

Similarly to ordinary QCD, we define singlet and non-singlet distributions

\beqa
 q_V(x, Q^2) &=& \sum_{i=1}^{n_f}\left( {q}_i(x,Q^2) - {\bar{q}}_i(x, Q^2)\right) , \nonumber \\
 \tilde{q}_V(x, Q^2) &=& \sum_{i=1}^{n_f}
\left( \tilde{q}_i(x, Q^2) - \tilde{\bar{q}}_i(x, Q^2)\right) \nonumber \\
 q^{+}(x, Q^2) &=& \sum_{i=1}^{n_f} \left({q}_i(x, Q^2) + \bar{q}_i(x, Q^2)\right) \nonumber \\
\tilde{q}^{+}(x, Q^2)&=& \sum_{i=1}^{n_f}\left( \tilde{q}_i(x,Q^2) + \tilde{\bar{q}}_i(x, Q^2)\right).
\eeqa

The evolution equations can be separated in two non-singlet sectors and a singlet one. 
The non-singlet are  

\beqa
Q^2 {d\over d Q^2} \left[\begin{array}{c}   q_V(x,Q^2)\\ \tilde{q}_V(x,Q^2)  \end{array} \right]
&=&{\alpha(Q^2)\over 2 \pi}P^{NS}_{SAP}\otimes 
 \left[\begin{array}{c}   q_V(x,Q^2)\\ \tilde{q}_V(x,Q^2)  \end{array} \right]
\label{susyns}
\eeqa

where the non singlet (NS) kernel is given by 
\beq
P^{NS}_{SAP}=\left[ \begin{array}{ll} 
         P_{qq} & P_{q\tilde{q}} \\
         P_{\tilde{q}q} & P_{\tilde{q}\tilde{q}} \end{array} \right]
\eeq

and the singlet, which mix $q_V$ and $\tilde{q}_V$ with the gluons and the gluinos are 

\beq
 Q^2 {d\over d Q^2}  \left[\begin{array}{c}   G(x,Q^2)\\ 
\lambda(x,Q^2) \\ q^+(x,Q^2) \\\tilde{q}^+(x,Q^2)  \end{array} \right]
=P^{S}_{ESAP}(x) \otimes
  \left[ \begin{array}{c}        
         G(x,Q^2)\\ \lambda(x,Q^2) \\ q^+(x,Q^2) \\ \tilde{q}^+(x,Q^2) \end{array} \right].
\eeq

where we have defined 

\beq
P^{S}_{ESAP}=\left[ \begin{array}{llll} 
         P_{GG} & P_{G \lambda} & P_{G q} &  P_{G \tilde{q}} \\
         P_{\lambda G} & P_{\lambda \lambda} & P_{\lambda q} & P_{\lambda \tilde{q}} \\
         P_{q G} & P_{q \lambda} & P_{q q} & P_{q s} \\
         P_{s G} & P_{s \lambda} & P_{\tilde{q} q} & P_{\tilde{q} \tilde{q}} 
         \end{array} \right]
\eeq

There are simple ways to calculate the kernel of the supersymmetric evolution by a simple extension of the usual methods. 
The changes are primarily due to color factors. There are also some
basic supersymmetric relations which have to be satisfied that will be
analized below. They are generally broken in the case of decoupling.  
We recall that the supersymmetric versions of the $\beta$ functions are given at 1-loop level by
\beqa
\beta_0^S &=& \frac{1}{3}\left(11 C_A - 2 n_f - 2 n_\lambda \right) 
\nonumber \\
\beta_0^{ES} &=& 3 C_A - \frac{1}{2}(n_L + n_R)
\eeqa
with $n_R=n_L=n_F=n_f/2$.
The running of the coupling is given by
\beq
{\alpha^{S,ES}(Q_0^2)\over 2 \pi}={2\over \beta_0^{S,ES}}
{ 1\over \ln (Q^2/\Lambda^2)}
\eeq

We use the ansatz discussed in \cite{CC1} which amounts to take
standard iterates of the convolution products in a large number to solve
the equations, implemented in the form of recursion relations fixed at
runtime. We start applying the method to the non-singlet case and then
move to the singlet. 
\section{Models of evolution}
We are going to discuss simple models of the supersymmetric 
evolution which incorporate several phases: the N=0 or regular QCD phase,
with fully decoupled superpartners, described by the usual DGLAP equations;
the N=1 phase, characterized by a decoupling of some of the superpartners 
(the squarks) from the remaining evolution, and, finally, the exact supersymmetric phase in which 
all the fields are effectively massless and are evolved simultaneously.
The last two phases can be obtained as intermediate and final stages of 
an evolution with a gluino mixing.

Beside the initial scale $Q_i$, at which we start the evolution, 
the beginning and the end of the intermediate phase (SAP) 
will be denoted by two additional scales, denoted by $Q_S$ and $Q_{ES}$. 
At some point, we will choose to vary the location of these two scales,  and analize the impact of the evolution according to all possible scenarios which are implied by these choices.

Some of these scenarios will be unrealistic, while others are more supported 
by the current experimental lower bounds on the masses of the superpartners. 
All the cases that will be discussed below have the objective to illustrate in a sufficiently detailed way the main features of the susy evolution. 

We should mention, if not obvious, 
that the intermediate and final phases in the evolution 
(evolved by the SAP and the ESAP respectively) 
are all affected by ``asymmetric'' boundary conditions, 
which are characterized by some densities of the superpartners set to zero at a given intermediate threshold.  
 
This reasoning, as explained in our previous work, is in line with current 
approaches to the analysis of the QCD evolution 
in which the parton densities can be generated by radiative collinear 
emissions from low scale distributions. 
We recall that the role of scaling violations and of the renormalization 
group all-together is, in this context, 
to simply dress light-cone matrix elements by logarithmic enhancements.    

It is convenient to introduce a general notation for all the kernels that 
will be considered. We embed both the singlet AP and SAP kernels into $4\times 4$ matrices 

\beq
P^{S}_{AP}=\left[ \begin{array}{llll} 
         P_{GG} & 0 & P_{G q} & 0 \\
         0 & 0 & 0 & 0 \\
         P_{q G} & 0 & P_{q q} & 0 \\
         0 & 0 & 0 & 0  
         \end{array} \right]
\eeq

\beq
P^{S}_{SAP}=\left[ \begin{array}{llll} 
         P_{GG} & P_{G \lambda} & P_{G q} & 0 \\
         P_{\lambda G} & P_{\lambda \lambda} & P_{\lambda q} & 0 \\
         P_{q G} & P_{q \lambda} & P_{q q} & 0 \\
          0 & 0 & 0 & 0
         \end{array} \right]
\eeq

while the non singlet DGLAP kernel is rewritten in a $2\times 2$ form

\beq
P^{NS}_{AP}=\left[ \begin{array}{ll} 
         P_{q q} & 0 \\
         0 & 0  
         \end{array} \right]
\eeq

For convenience we introduce the following notations. A region in the RG evolution 
- regular, partially decoupled or exact supersymmetric - 
is described by an array $(Q_i,Q_f)$, where $Q_i$ is 
the initial evolution scale and $Q_f$ the final evolution scale. The type of 
evolution (AP,SAP,ESAP) is indicated by a corresponding suffix. For instance 
$(2,100)_{AP}$ denotes a regular DGLAP evolution with initial scale $Q_i=2$ GeV and a 
final scale of $Q_f=100$ GeV. Similarly, an evolution of the form $(2,100)_{AP}-(100,400)_{SAP}-(400,1000)_{ESAP}$ describes a matching of all the three evolutions 
at the intermediate scales $Q_{f,AP}=Q_{i,SAP}=100$ GeV and $Q_{f,SAP}=Q_{i,ESAP}=400$ GeV. The masses of the intermediate scales are denoted by an array with 2 entries 
$(m_{2\lambda},m_{2 \tilde{q}})$. In the example presented above, 
these two scales are the matching scales for the SAP and ESAP evolutions $Q_{i,SAP}=m_{2\lambda}$ and $Q_{i,ESAP}=m_{\tilde{q}}$. 
 As usual, we adopt a step approximation in the running, 
according to which we step into a new region right after crossing the corresponding 
mass threshold. 
We recall that we use a general logarithmic ansatz for the structure of the logarithmic
contributions in which the momentum dependence of the scaling violations
is parametrized by the actual running coupling. In \cite{CC1} we have
elaborated on this point in some detail. We set the expansion

\beq
{A}_0^{f}(x) =  \sum_{n=0}^{\infty}{{A}^{f}_{n}(x)\over n!} 
\log^n\left({\alpha(Q)\over \alpha(Q_0)}\right). \nonumber  \\
\label{grossi}
\eeq
and invoke an equality between logarithmic powers at the left-hand-side
and at the right-hand-side of the equations to get recursion relations
for the coefficients ${A^f}_n(x)$.

TO be specific, let's consider the most general sequence of evolutions (AP-SAP-ESAP) described by the arrays 
$(Q_i,Q_f)_{AP}-(Q_i-Q_f)_{SAP}-(Q_i,Q_f)_{ESAP}$, with $Q_{f,AP}=Q_{i,SAP}=m_{2\lambda}$ 
and $Q_{f,SAP}=Q_{i,ESAP}=m_{\tilde{q}}$. 
In this (general) case the solution is built by sewing the three regions as  

\beqa
\left[ \begin{array}{c} 
         q_V(x,Q^2)\\
         \tilde{Q}_V(x,Q^2)        \\	
         \end{array} \right] &=& \left[ \begin{array}{c} 
         q_V(x,Q_0^2)\\
         0        \\	
         \end{array} \right] + \int_{Q_0^2}^{m_{2\lambda}^2} d\, \log\, Q^2\, 
P_{AP}^{NS}(x,\alpha(Q^2))\otimes \left[ \begin{array}{c} 
         q_V(x,Q^2)\\
         0        \\	
         \end{array} \right] \nonumber \\
&+& \int_{m_{2\lambda}^2}^{m_{ \tilde{q}}^2} d\, \log Q^2 
P_{SAP}^{NS}(x,\alpha^S(Q^2))\otimes \left[ \begin{array}{c} 
         q_V(x,Q^2)\\
         0        \\	
         \end{array} \right] \nonumber \\
&+& \int_{m_{2\lambda}^2}^{Q_f^2} d\, \log Q^2\, 
P_{ESAP}^{NS}(x,\alpha^{ES}(Q^2))\otimes \left[ \begin{array}{c} 
         q_V(x,Q^2)\\
         \tilde{q}^V(x,Q^2)        \\	
         \end{array} \right] \nonumber \\
\eeqa
in the non singlet and

\beqa
\left[ \begin{array}{c} 
         G(x,Q_f^2)\\
	 \lambda(x,Q^2)\\
	 q^+(x,Q^2)\\
         \tilde{q}^+(x,Q^2)\\	
         \end{array} \right] &=& \left[ \begin{array}{c} 
         G(x,Q_f^2)\\
	 0\\
	  q^+(x,Q_f^2) \\
         0\\	
         \end{array} \right]  + \int_{Q_0^2}^{m_{2\lambda}^2} d\, \log\, Q^2\, 
P_{AP}^{NS}(x,\alpha(Q^2))\otimes 
\left[ \begin{array}{c} 
         G(x,Q_f^2)\\
	  0\\
	 q^+(x,Q^2)\\
         0\\	
         \end{array} \right] \nonumber \\
&+& \int_{m_{2\lambda}^2}^{m_{ \tilde{q}}^2} d\, \log Q^2 
P_{SAP}^{NS}(x,\alpha^S(Q^2))\otimes 
\left[ \begin{array}{c} 
         G(x,Q_f^2)\\
	 \lambda(x,Q^2)\\
	 q^+(x,Q^2)\\
           0\\	
         \end{array} \right] \nonumber \\
&+& \int_{m_{2\lambda}^2}^{Q_f^2} d\, \log Q^2\, 
P_{ESAP}^{NS}(x,\alpha^{ES}(Q^2))\otimes
\left[ \begin{array}{c} 
         G(x,Q_f^2)\\
	 \lambda(x,Q^2)\\
	 q^+(x,Q^2)\\
           \tilde{q}^+(x,Q^2)\\
         \end{array} \right] \nonumber \\
\eeqa
for the singlet solution. 
The zero entries in the arrays for some of the distributions are due 
to the boundary conditions, since all the supersymmetric partners are generated, in this model, 
by the evolution. The general structure of the algorithms that solves these equations 
is summarized below. We start from the non singlet. We define 
$q^{NS}(x,Q^2)=\left(q_V(x,Q^2),\tilde{q}_V(x,Q^2)\right)^T$  and 
$A_n(x)=\left(A^{q_V}_n,A^{\tilde{q}_V}_n\right)^T$ and introduce the ansatz

\beq
q^{NS}(x,Q^2)=\sum_{n=0}^{n_0}\,{A_n(x)\over n!} \log^n
\left(\frac{\alpha(Q^2)}{\alpha{(Q_0^2)}}\right),
\eeq
where $n_0$ is an integer at which we stop the iteration. 
Usually ranges between 30 and 40.
The first coefficient of the recursion is determined by the initial condition
\beq
A_0(x) = U(x)\otimes q^{NS}(x,Q_0^2), 
\eeq
where
\beqa
U_1(x) &\equiv & 
\left[ \begin{array}{ll} 
         \delta(1-x) & 0\\
	  0 & 0\\	
         \end{array} \right] \nonumber \\
U_{1\,2}(x) &\equiv & 
\left[ \begin{array}{ll} 
         \delta(1-x) & 0\\
	  0 & \delta(1-x)\\	
         \end{array} \right]. 
\eeqa

The recursion relations are given by 

\beq
A_{n+1}(x)= -\frac{2}{\beta_0}P^{NS}_{AP}\otimes A_n(x)
\eeq
The solution in the first (DGLAP) region at the first macthing scale 
$m_{2\lambda}$ is given by 
\beq
q^{NS}(x,m_{2\lambda})=\sum_{n=0}^{n_0}\frac{A^{AP}_n}{n!}\log^n
\left(\frac{\alpha(m_{2\lambda}^2)}{\alpha(Q_0^2)}\right).
\eeq
At the second stage the (partial) supersymmetric coefficients are given by 
(S is a short form of $SAP$)

\beqa
A^{S,NS}_0(x) &=& U(x)\otimes q(x,m_{2\lambda^2}) \nonumber \\
A^{S,NS}_{n+1}(x) &=& -\frac{2}{\beta^S_0} P^{NS}_{S}(x)\otimes A^{S,NS}_n(x).\nonumber \\
\eeqa
We construct  the boundary condition for the next stage of the evolution using the intermediate solution 
\beq
q(x,Q^2)=\sum_{n=0}^{n_0}\frac{A^S_n(x)}{n!}
\left(\frac{\alpha(m_{Q^2})}{\alpha(m_{2\lambda}^2)}\right).
\eeq
evaluated at the next threshold $m_{\tilde{q}}$. 
\beq
q(x,\tilde{q}^2)=\sum_{n=0}^{n_0}\frac{A^S_n(x)}{n!}
\log\left(\frac{\alpha({Q^2})}{\alpha(m_{2\lambda}^2)}\right).
\eeq
The final solution is constructed using the recursion relations 
\beqa
A^{ES}_0(x) &=& U(x)\otimes q(x,m_{ \tilde{q}}) \nonumber \\
A^{ES}_{n+1}(x) &=& -\frac{2}{\beta^{ES}}P^{NS}_{ES}(x)\otimes A^{ES}_n(x) \nonumber \\
\eeqa 

The final solution is written as 
\beqa
q^{NS}(x,Q^2) &=& q^{NS}(x,Q_0^2) + \sum_{n=1}^{n_0}
\frac{A_n(x)}{n!}\log\left(\frac{\alpha(m_{2\lambda}^2)}{\alpha(Q_0^2)}\right)
+ 
\sum_{n=1}^{n_0}\frac{A^{S}_n(x)}{n!}\log\left(\frac{\alpha^{S}(m_{\tilde{q}}^2)}
{\alpha^{S}(m_{2\lambda}^2)}\right) \nonumber \\
& + & \sum_{n=1}^{n_0}\frac{A^{ES}_n(x)}{n!}\log\left(\frac{\alpha^{ES}(Q_f^2)}
{\alpha^{ES}(m_{\tilde{q}}^2)}\right) \nonumber \\ 
\eeqa

\section{Solving the non singlet equations}
In the case of exact supersymmetry, the arguments and the strategy
presented in \cite{CC1} simplify, since we neglect all the
intermediate scales and the evolution -starting from the lowest scale-
is assumed to be supersymmetric. The boundary contions at the start of the evolution, however, 
are not. We implement the algorithm as it has been formulated in our previous work, 
with due modifications given the different structure of the evolution kernels. 

The recursion relations needed in the implementation are defined in
terms of recursive coefficients ${A^f}_{n}(x)$ with
$n=0,1,2,..$.

This equation can be made explicit by separating the more singulare terms 
of the recursion relations from the rest

\beqa
A_{n+1}^{q^v}(x) &=& -\frac{4}{\beta_0^S}C_F\int_{x}^1\frac{dy}{y}
\frac{y\,\, A_n^{q^v}(y) - x\,\,A_n^{q^v}(x)}{y- x}
+\frac{2}{\beta_0^S}C_F\int_x^1\frac{dy}{y}\left( 1+ z\right)A_n^{q^v}(y)
\nonumber \\
&& - \frac{2}{\beta_0^S}C_F A_n^{q^v}(x) -\frac{2}{\beta_0^S}C_F \int_x^1 
\frac{dy}{y}A_n^{\tilde{q}^v}(y)  -\frac{4}{\beta_0^S}C_F \log(1-x)A_n^{q^v}(x)
\nonumber \\
\eeqa

\beqa
{A}^{\tilde{q}^v}(x)&=&-\frac{2}{\beta_0^S}C_F\int_x^1
\frac{dy}{y}A_n^{q^v}(y) +
\frac{2}{\beta_0^S}C_F\int_x^1\frac{dy}{y}
\left(1 +z\right)A_n^{\tilde{q}^v}(y) -\frac{2}{\beta_0^S}C_FA_n^{\tilde{q}^v}(x)
\nonumber \\
& &- \frac{4}{\beta_0^S} C_F \int_{x}^1\frac{dy}{y}
\frac{y\,\, A_n^{\tilde{q}^v}(y) - x\,\,A_n^{\tilde{q}^v}(x)}{y- x}
 -\frac{4}{\beta_0^S} C_F \log(1-x)A_n^{\tilde{q}^v}(x)
\nonumber \\
\eeqa

\section{The Singlet Equation}
The singlet equations are treated in a similar way. 
The edge-point contributions ($x=1$) appear to be different from the usual QCD
expressions for those splitting functions that are part of the regular
QCD evolution $P_{gg}$ and $P_{qq}$ due to the supersymmetry relations

\beq
P_{gg} + P_{\lambda g} =  P_{g \lambda} + P_{\lambda \lambda} \\
\label{firstrelation}
\eeq
and by 
\beqa
P_{q g} + P_{\lambda q} & = & P_{g s} + P_{\lambda s}\nonumber \\
P_{q g} + P_{s g}  & = & P_{q \lambda} + P_{s\lambda}\nonumber \\
P_{qq} + P_{s q} & = & P_{q s} + P_{s s}.
\eeqa
The recursion relations, in this case, become 

\beqa
{A}_{n+1}^{q+} & = & 
-\frac{4 C_F}{\beta_0^S} \int_x^1 \frac{dy}{y}
\frac{ y {A}_n^{q +}(y) - x {A}_n^{q +}(x)}{ y - x}
-\frac{2}{\beta_0^S}C_F\int_x^1\frac{dy}{y}(1+z)A_n^{q^+}(y)\nonumber \\
& - & \frac{4 C_F}{\beta_0^S}\log(1-x) {A}_n^{q +}(x) 
-{2\over{\beta_0^S}}C_F A_n^{q^+}(x)\nonumber \\
& & -\frac{2}{\beta_0^S}T_R \int_x^1\frac{dy}{y}\left( 2 z^2 - 2 z +1\right)A_n^g(y)
\nonumber \\
& & -\frac{2}{\beta_0^S}T_R\int_x^1\frac{dy}{y}( 1-z)A_n^\lambda(y)
-\frac{2}{\beta_0^S}C_F \int_x^1\frac{dy}{y}\left(\frac{2}{z}-2\right)A_n^{\tilde{q}^v}(y)
\eeqa

\beqa
{A}_{n +1}^\lambda(x) & = & -\frac{2}{\beta_0^S}C_F
\int_x^1\frac{dy}{y}\left( 1- z\right) {A}_n^{q^+} 
-4 \frac{C_A}{\beta_0^S}\int_x^1\frac{dy}{y} 
\frac{ y {A}_n^\lambda(y) - x{A}_n^\lambda(x)}{y -x} \nonumber \\
&& -\frac{4}{\beta_0^S}C_A\log(1-x)A_n^\lambda(x)
+ \frac{2}{\beta_0^S} C_A \int_x^1 \frac{dy}{y}(1 + z) A_n^\lambda(y) 
-\frac{2}{\beta_0^S} \left(\frac{3}{2}- \frac{T_R}{2}\right)A_n^\lambda(x)\nonumber \\
&& -\frac{2}{\beta_0^S}C_A\int_x^1\frac{dy}{y}\left( z^2 +(1-z)^2\right)A_n^g(y)
-\frac{2}{\beta_0^S}C_F\int_x^1\frac{dy}{y}A_n^{\tilde{q}^+} \nonumber \\
\eeqa

\beqa
{A}_{n+1}^g &=& 
-\frac{2}{\beta_0^S}C_F\int_x^1\frac{dy}{y}\left(\frac{2}{z}-2 +z\right)A_n^{q^+}(y)
-\frac{4}{\beta_0^S} C_A\int_x^1\frac{dy}{y}\frac{y A_n^g(y) -x A_n^g(x)}{y-x}
\nonumber \\
& & - \frac{4}{\beta_0^S} C_A \log(1-x) A_n^g(x)
+ \frac{2}{\beta_0^S} C_A\int_x^1\frac{dy}{y}(1+z)A_n^g(y)
\nonumber \\
&& -\frac{2}{\beta_0^S} C_A\int_x^1\frac{dy}{y}\left(\frac{2}{z} +z -2\right)
A_n^g(y) + \frac{2}{\beta_0^S} C_A\int_x^1\frac{dy}{y}\left(z^2 + (1-z)^2\right)A_n^g(y)
\nonumber \\
&& -\frac{2}{\beta_0^S} \left(\frac{3}{2} C_A - \frac{T_R}{2}\right)A_n^g(x)
- \frac{2}{\beta_0^S}C_A \int_x^1\frac{dy}{y}\left( \frac{2}{z}+z -2\right) A_n^\lambda(y)
\nonumber \\
&& -\frac{2}{\beta_0^S}C_F \int_x^1 \frac{dy}{y}\left(\frac{2}{z}-2\right)A_n^{\tilde{q}^+}(y) 
\nonumber \\
\eeqa

\beqa
A_{n+1}^{\tilde{q}^+}&=&-\frac{2}{\beta_0^S}C_F \int_x^1\frac{dy}{y} z A_n^{q^+}(y)
\nonumber \\
&& - \frac{2}{\beta_0^S}T_R\int_x^1\frac{dy}{y}\left(1-z^2 -(1-z)^2\right)A_n^g(y)
-\frac{2}{\beta_0^S}T_R \int_x^1\frac{dy}{y}z A_n^{\lambda}(y)\nonumber \\
&& -\frac{4}{\beta_0^s}C_F \int_x^1\frac{dy}{y}
\frac{y A_n^{\tilde{q}^+}- x A_n^{\tilde{q}^+}(x)}{y-x}\nonumber \\
& & -\frac{4}{\beta_0^S} C_F\log(1-x) A_n^{\tilde{q}^+} + \frac{4}{\beta_0^S} C_F\int_x^1\frac{dy}{y}A_n^{\tilde{q}^+}
(y) -\frac{2}{\beta_0^S}C_F A_n^{\tilde{q}^+}(x)\nonumber \\
\eeqa

\section{AP-ESAP evolution} 
As a starting point, we discuss a model of the evolution in which supersymmetry is switched-on right above the ordinary QCD evolution, without introducing 
an intermediate region of partial decoupling. This is equivalent to choose
$m_{\lambda}=m_{\tilde q}$.
This example may serve 
as an illustration of the impact of a full (or exact) supersymmetric 
evolution on top of the regular QCD evolution.
Initial conditions 
for the ESAP evolution are characterized by vanishing densities 
for all the superpartners at the scale $Q_{ES}$ where the ESAP evolution starts, 
and by (leading order) evolved AP distributions of ordinary QCD.  
We plot only distributions summed over all the flavours of quarks and 
squarks such as $x q^{+}(x)$, 
$x \tilde{q}^{+}(x)$ and we fix the number of flavours $n_f=4$.

The valence quark distributions $q_V(x,Q_0^2)$ and gluon distributions $G(x,Q_0^2)$ at the input scale $Q_0$, taken from the 
CTEQ3M parametrization \cite{cteq}
\beqn
q(x)&=&A_0 x^{A_1}(1-x)^{A_2}(1 + A_3 x^{A_4}).
\eeqn

Specifically

\beqn
x u_V(x)&=&1.37x^{0.497}(1-x)^{3.74}[1 + 6.25 x^{0.880}] \nonumber \\
x d_V(x)&=&0.801 x^{0.497}(1-x)^{4.19}[1 + 1.69 x^{0.375}] \nonumber \\
x G(x)&=&0.738 x^{-0.286}(1-x)^{5.31}[1 + 7.30 x] \nonumber \\
x(\frac{\bar{u}(x) +\bar{d}(x)}{2}) &=& 0.547 x^{-0.286}(1-x)^{8.34}[1 +
17.5 x]\nonumber \\
x s(x)&=& 0.5x(\frac{\bar{u}(x) +\bar{d}(x)}{2})
\eeqa
and a vanishing anti-strange contribution at the input. Fig.1 shows the shapes of the initial CTEQ distributions at an initial scale 
of $Q_i=2$ GeV. 
%%%%%%%%%%%%%%%%%%%%%%%%%%%%%%%%%%%%%
\section{AP-ESAP Evolution}
\begin{figure}[thb]
\centerline{\includegraphics[angle=-90,width=.9\textwidth]{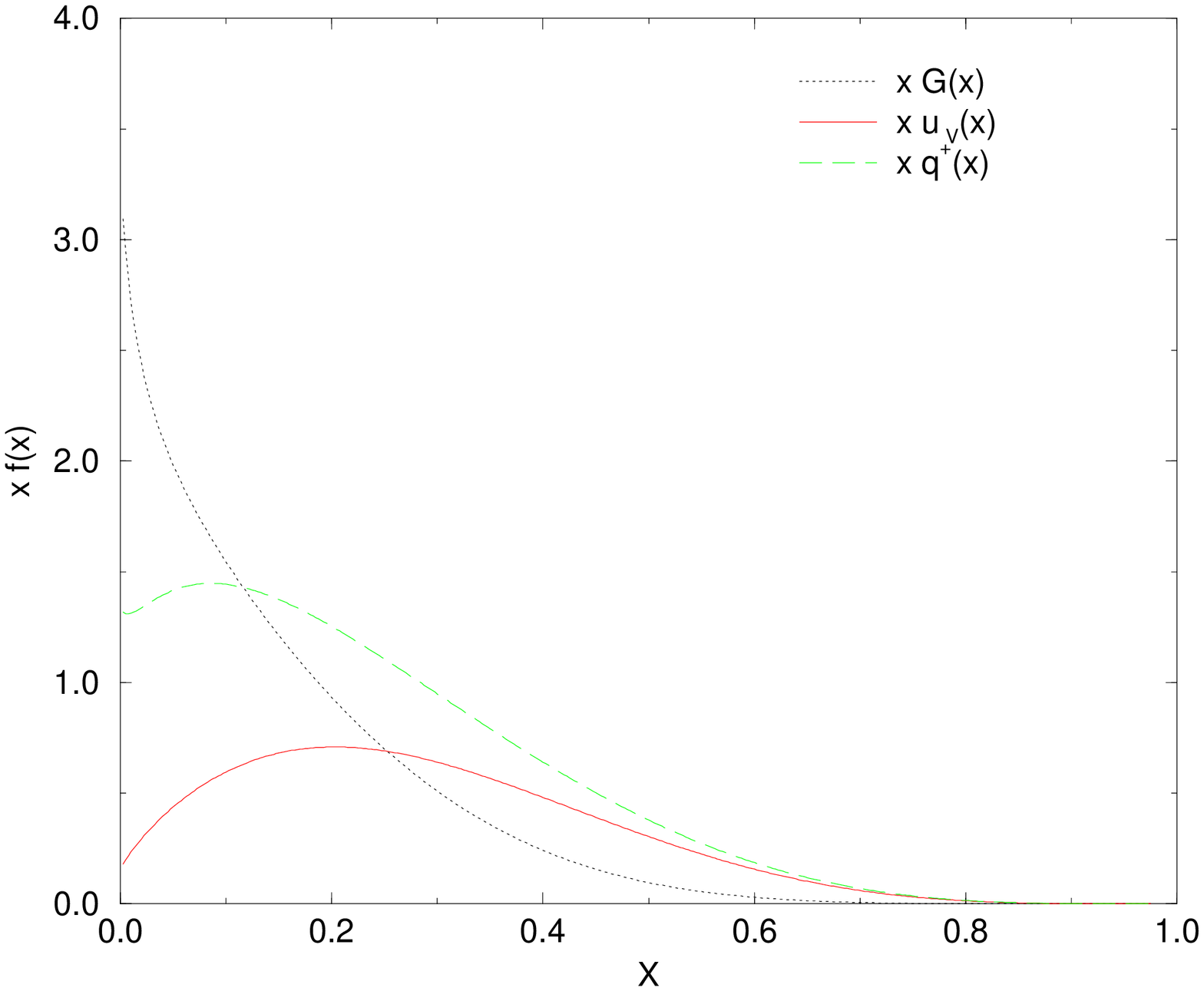}}
\caption{{initial CTEQ3M distributions used for a regular AP, 
with $Q_f=3$ GeV.}}
\label{quark1}
\centerline{\includegraphics[angle=-90,width=.9\textwidth]{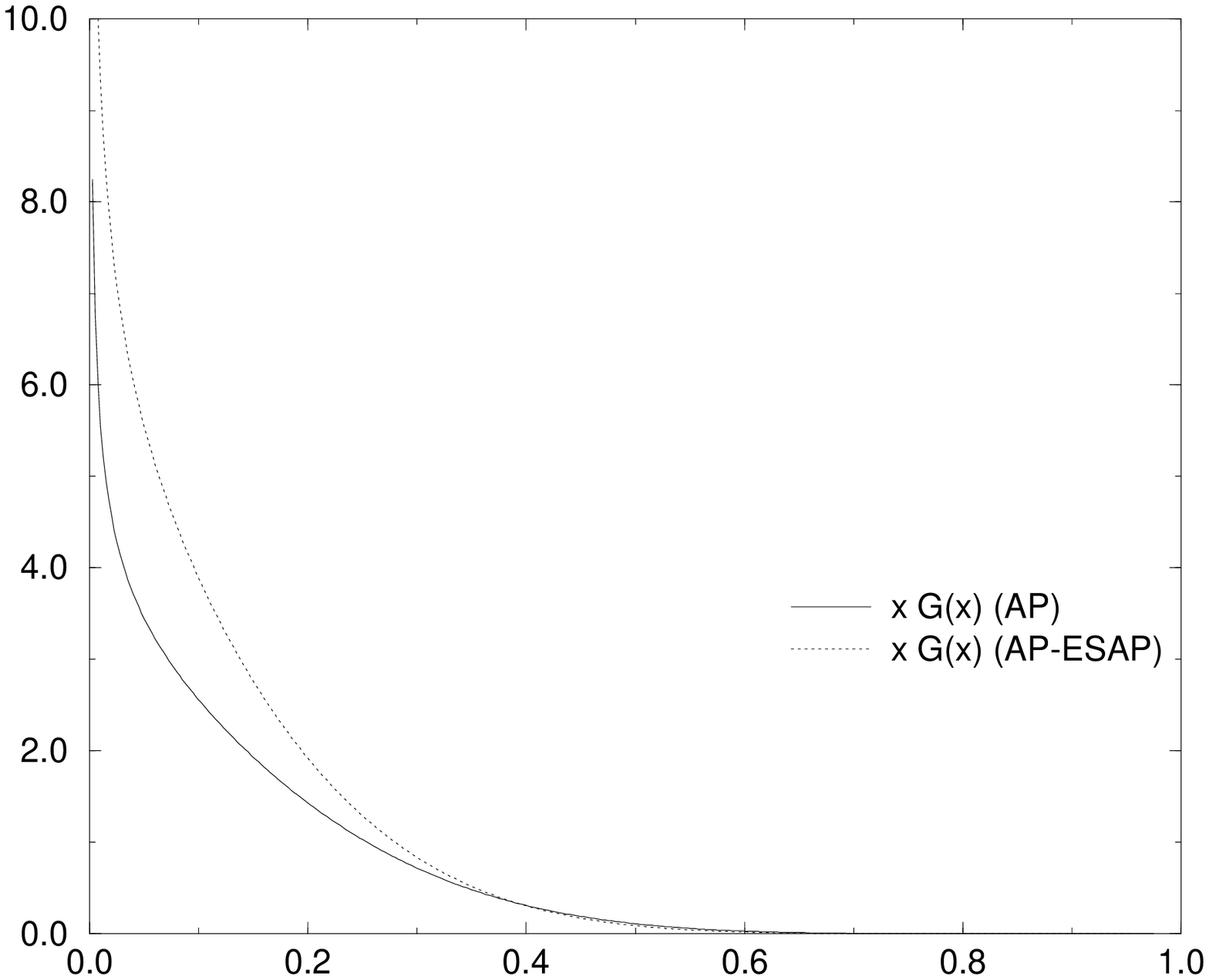}}
\caption{{$x G(x)$ for a regular AP versus an ESAP evolution, 
with $m_{2\lambda}=20$ GeV. The final evolution scale is $Q_f=500$ GeV }}
\label{quark2}
\end{figure}

\begin{figure}
\centerline{\includegraphics[angle=-90,width=.9\textwidth]{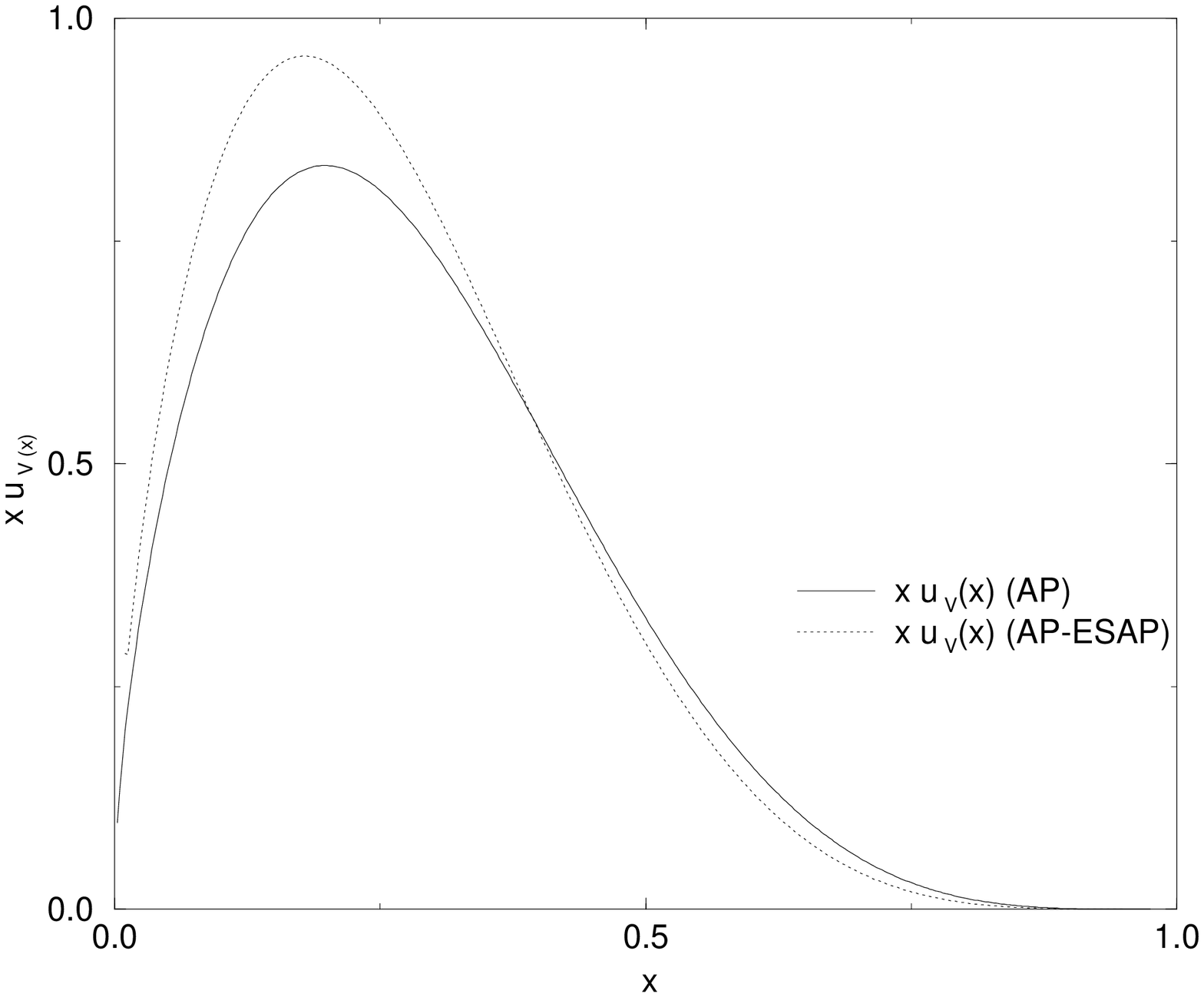}}
\caption{{$x u_V(x)$ for a regular AP versus an AP-ESAP evolution, 
with $m_{2\lambda}=20$ GeV. The final evolution scale is $Q_f=500$ GeV }}
\label{quark3}
\centerline{\includegraphics[angle=-90,width=.9\textwidth]{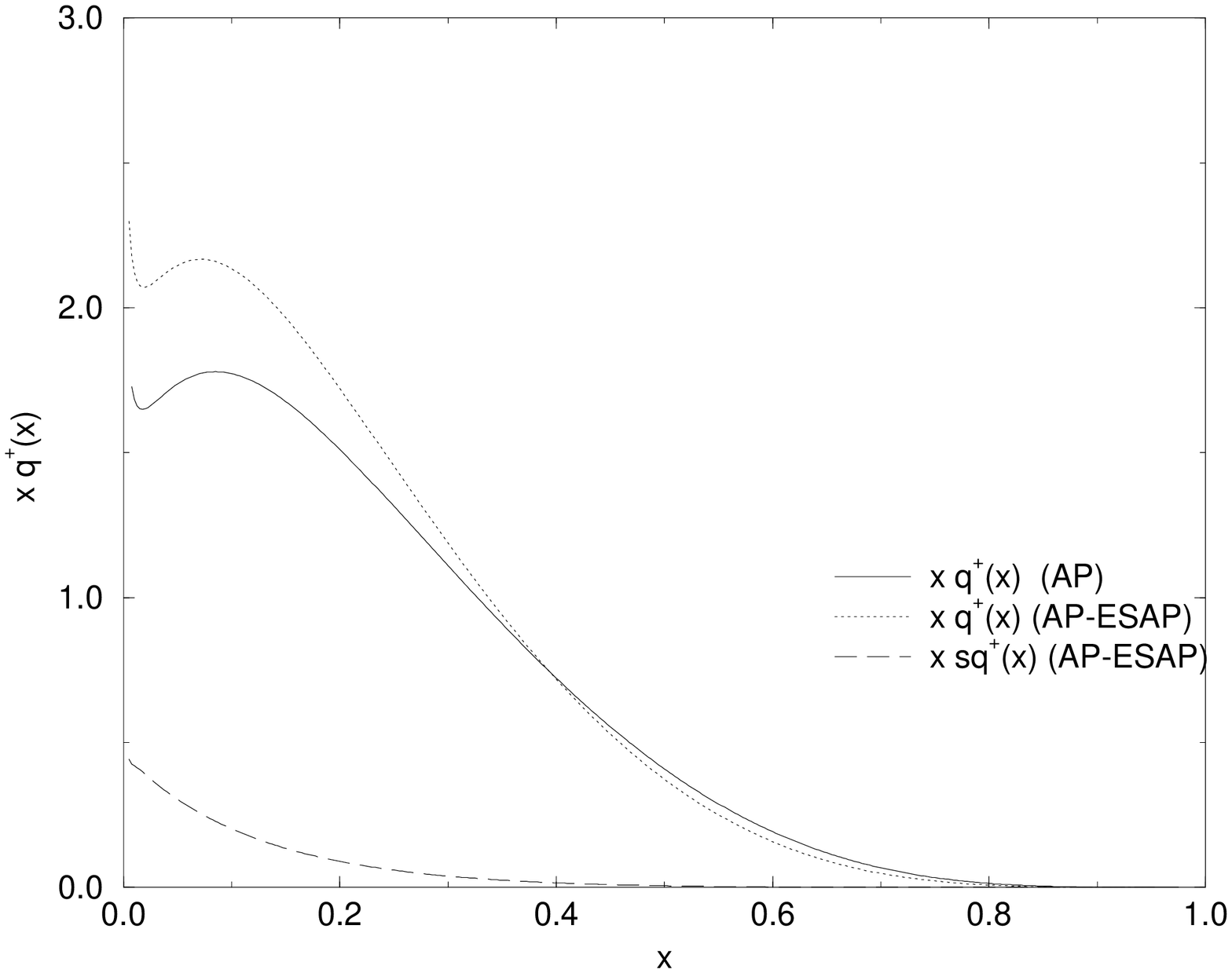}}
\caption{{$x q^+(x)$ and $x \tilde{q}^+(x)$ for a regular AP versus an AP-ESAP evolution, 
with $m_{2\lambda}=20$ GeV. The final evolution scale is $Q_f=500$ GeV }}
\label{quark4}
\end{figure}

We show in Fig.~\ref{quark1} the shape of the distributions 
at the lower scale of 3 GeV. They are the non-singlet $x u_V(x)$, 
the singlet combination $ x q^+(x)$ and the gluon $x G(x)$. 
We will be using these initial shapes all along in our analysis. 
In Figs.~\ref{quark2},\ref{quark3} and \ref{quark4} we show comparisons between the AP evolution and the combined AP-ESAP evolution. As we have already mentioned, 
this model of the evolution is realistic if all the susy partners are close in mass. 
We start focusing our attention on the regular QCD distributions, now evolved in an 
mixed (regular and supersymmetric) setting. 
We take the 
matching parameter $m_{ 2\lambda}$ to be of 20 GeV, corresponding 
to a light gluino (and to a light scalar quark). The final evolution scale is 
$Q_f=500$ GeV. In Fig.~\ref{quark2} we show the evolution 
of the gluon density in the AP case versus the AP-ESAP case. 
The initial and final scales of the AP evolution are taken to be the same 
($Q_0=2$ GeV and $Q_f=$500 GeV). 
Scaling violations are sizeable both for gluons, for valence quarks 
(see Fig.\ref{quark4}) and for singlet quarks. As shown in Fig.~\ref{quark4} 
the singlet squark density becomes significant at smaller x. 
In this figure we also compare in size this distribution to the singlet quark one. 
This distribution is down approximately by a factor of 10 compared to that of 
regular quarks. The decrease at larger-x of this supersymmetric 
distribution is also much faster.  

\begin{figure}
\centerline{\includegraphics[angle=-90,width=.9\textwidth]{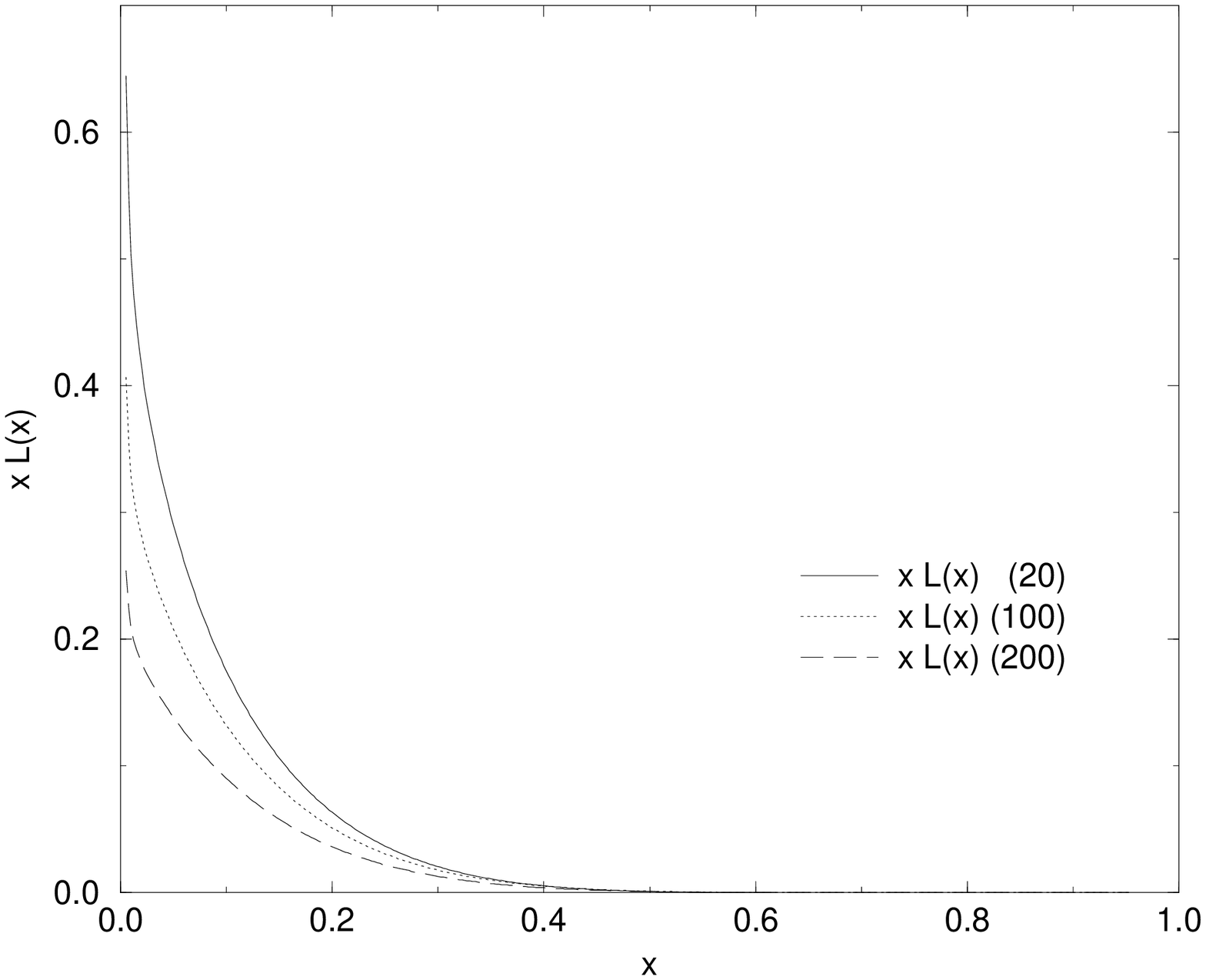}}
\caption{{$x \lambda(x)$ for the AP-ESAP evolution, 
with $m_{\tilde{q}}=20,100,200$ GeV respectively. The final evolution 
scale is $Q_f=500$ GeV }}
\label{quark5}
\centerline{\includegraphics[angle=-90,width=.9\textwidth]{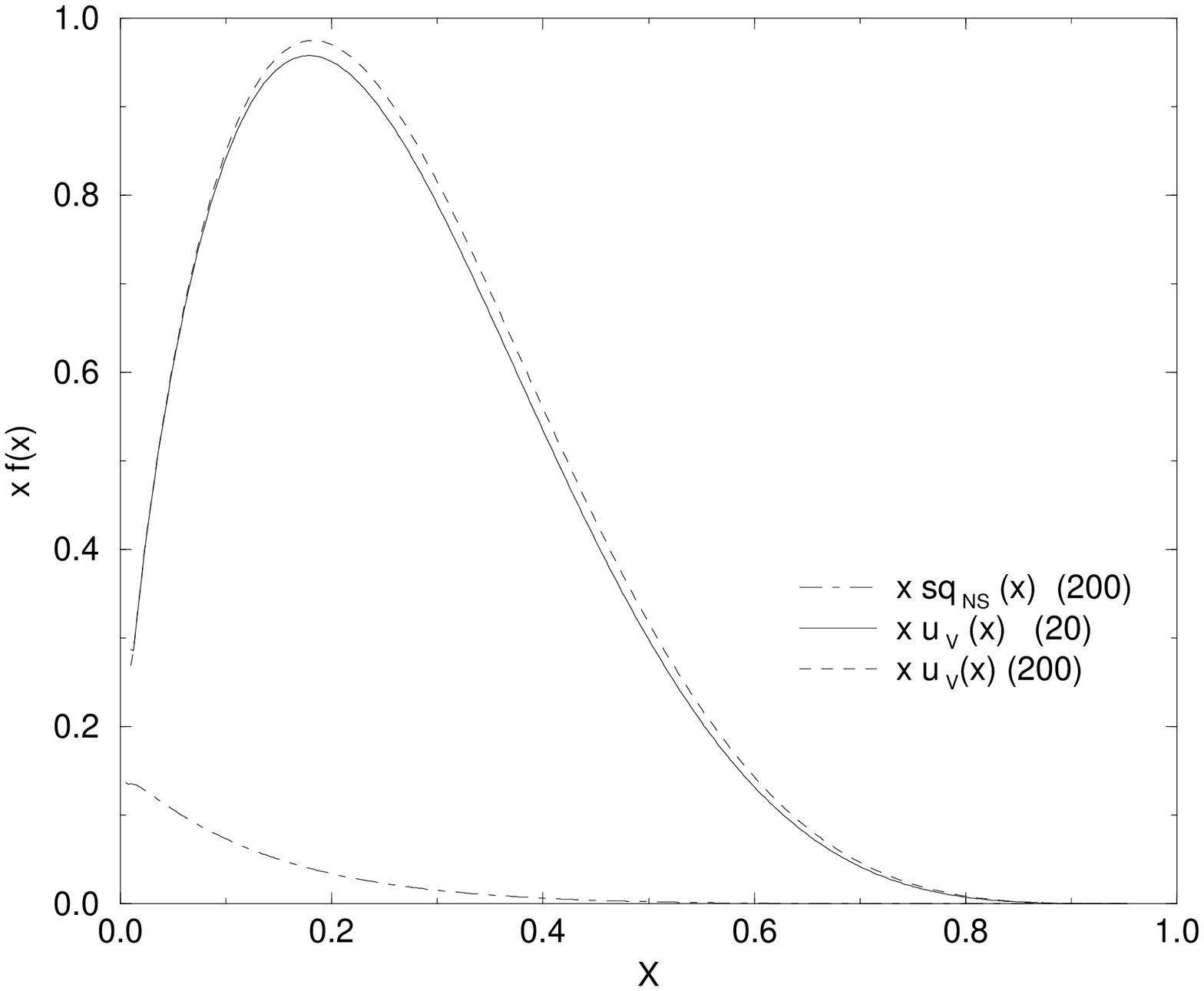}}
\caption{{ Non singlet distributions for the AP-ESAP evolution, 
with $m_{\tilde{q}}=20,200$ GeV for $x u_V(x)$ and for 
the non-singlet distribution of scalar quarks 
($m_{\tilde{q}}=200$ GeV). The final evolution scale is $Q_f=500$ GeV }}
\label{quark6}
\end{figure}

\begin{figure}
\centerline{\includegraphics[angle=-90,width=.9\textwidth]{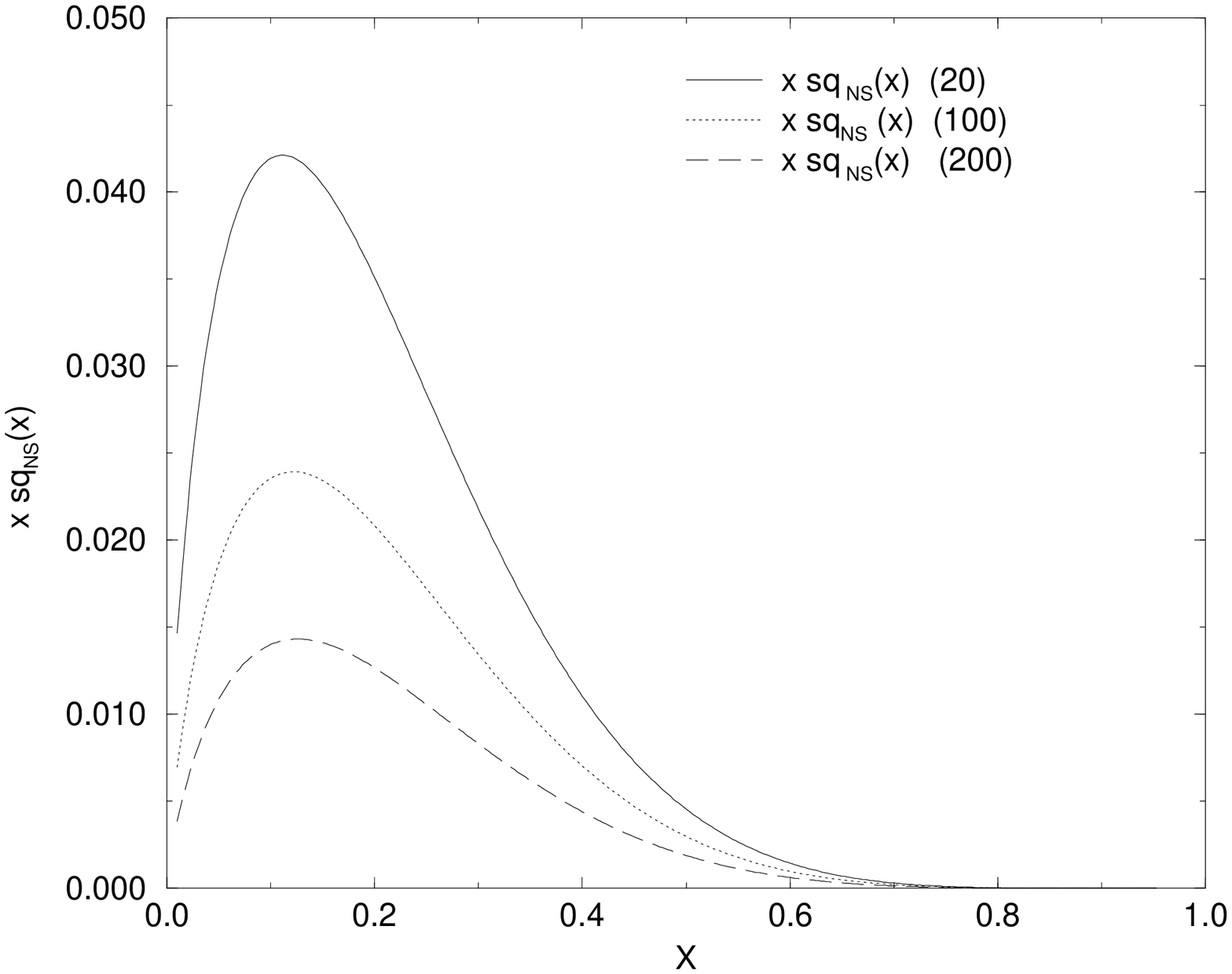}}
\caption{{ x $\tilde{q}^-(x)$ in the AP-ESAP evolution, 
with $m_{\tilde{q}}=20,100, 200$ GeV. The final evolution scale is $Q_f=500$ GeV }}
\label{quark8}
\centerline{\includegraphics[angle=-90,width=.9\textwidth]{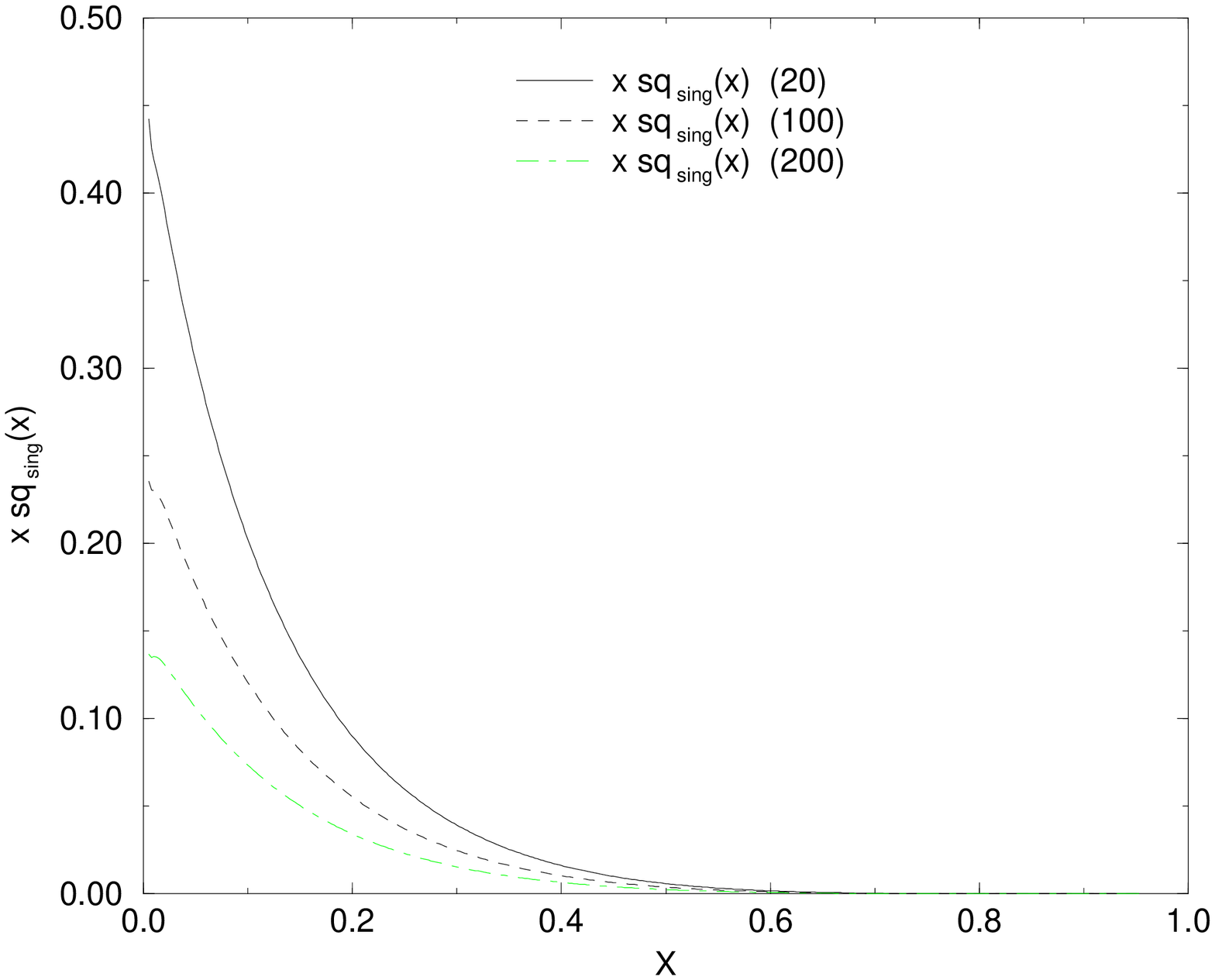}}
\caption{{ x $\tilde{q}^+(x)$ in the AP-ESAP evolution, 
with $m_{\tilde{q}}=20,100, 200$ GeV. The final evolution scale is $Q_f=500$ GeV }}
\label{quark9}
\end{figure}

A similar pattern is shown in Fig.~\ref{quark5}, in this case for the gluino 
density. The figure shows the dependence of the density on the 
matching scale $m_{\tilde{q}}$. The gluino distributions grows sharply at small x. 
As expected, as we increase the mass of the scalar quark the gluino distribution flattens. 
In Fig.~\ref{quark6}  we compare the valence quark distribution for 
two values of the scalar quark mass in the evolution 
($m_{\tilde{q}}=20$ GeV and 200 GeV respectively). 
The final evolution scale is fixed to 500 GeV. In the same plot we show for comparison the non singlet scalar quark distribution 
(for a mass $m_{\tilde{q}}=200$ GeV). 
The variation in shape of this distribution 
due to the presence of a supersymmetric threshold in the evolution is comparable to the typical scaling violations of ordinary QCD induced by a change of the factorization scale. 
In Fig.~\ref{quark8} we plot the non-singlet squark distribution (denoted as $sq_{ns}$) 
for a varying squark mass. The distribution gets lowered drastically 
as the mass of the scalar quark increases. 
The shape of this distribution, which is generated 
radiatively -starting from an initial scale $m_{\tilde{q}}$- 
is similar to the usual non singlet quark distribution, 
but rescaled by a factor approximately estimated to be of $5/100$. Fig.~\ref{quark9} shows the singlet quark 
density for 3 values of $m_{\tilde{q}}$. The distributions are more pronounced at smaller x values and become lowered as the scalar quark mass increases. This is expected, since, for a given 
supersymmetric evolution interval $(Q_{int},Q_f)_{ESAP}$, 
the supersymmetric interval gets smaller as $Q_{int}\to Q_f$ 
and therefore, the logarithmic enhancements are reduced. 

\begin{figure}
\centerline{\includegraphics[angle=-90,width=.9\textwidth]{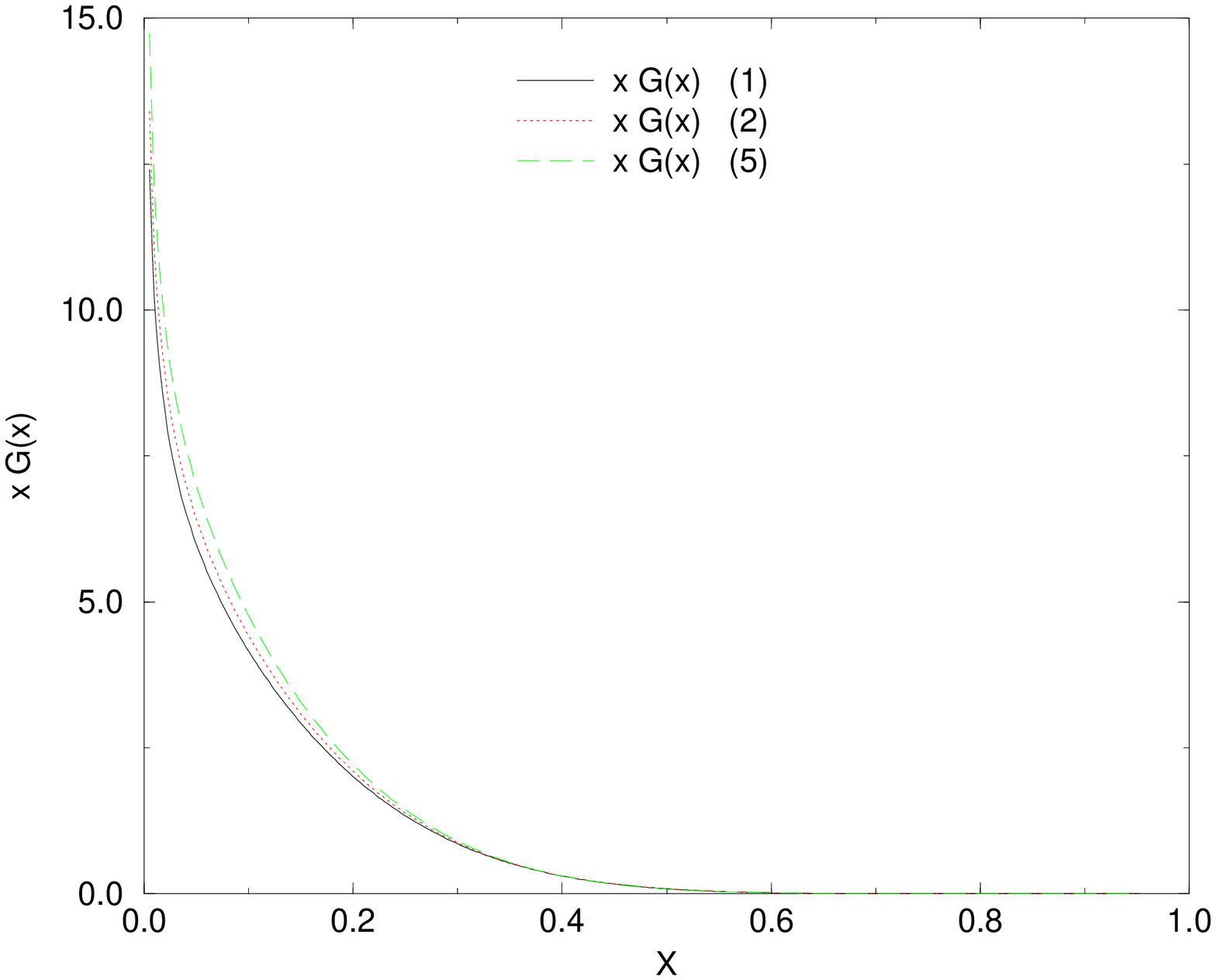}}
\caption{{ $x G(x)$ in the AP-ESAP evolution, 
with a squark mass $m_{\tilde{q}}=20$ GeV. The final evolution scales are 1,2, and 5 TeV respectively }}
\label{quark10}
\centerline{\includegraphics[angle=-90,width=.9\textwidth]{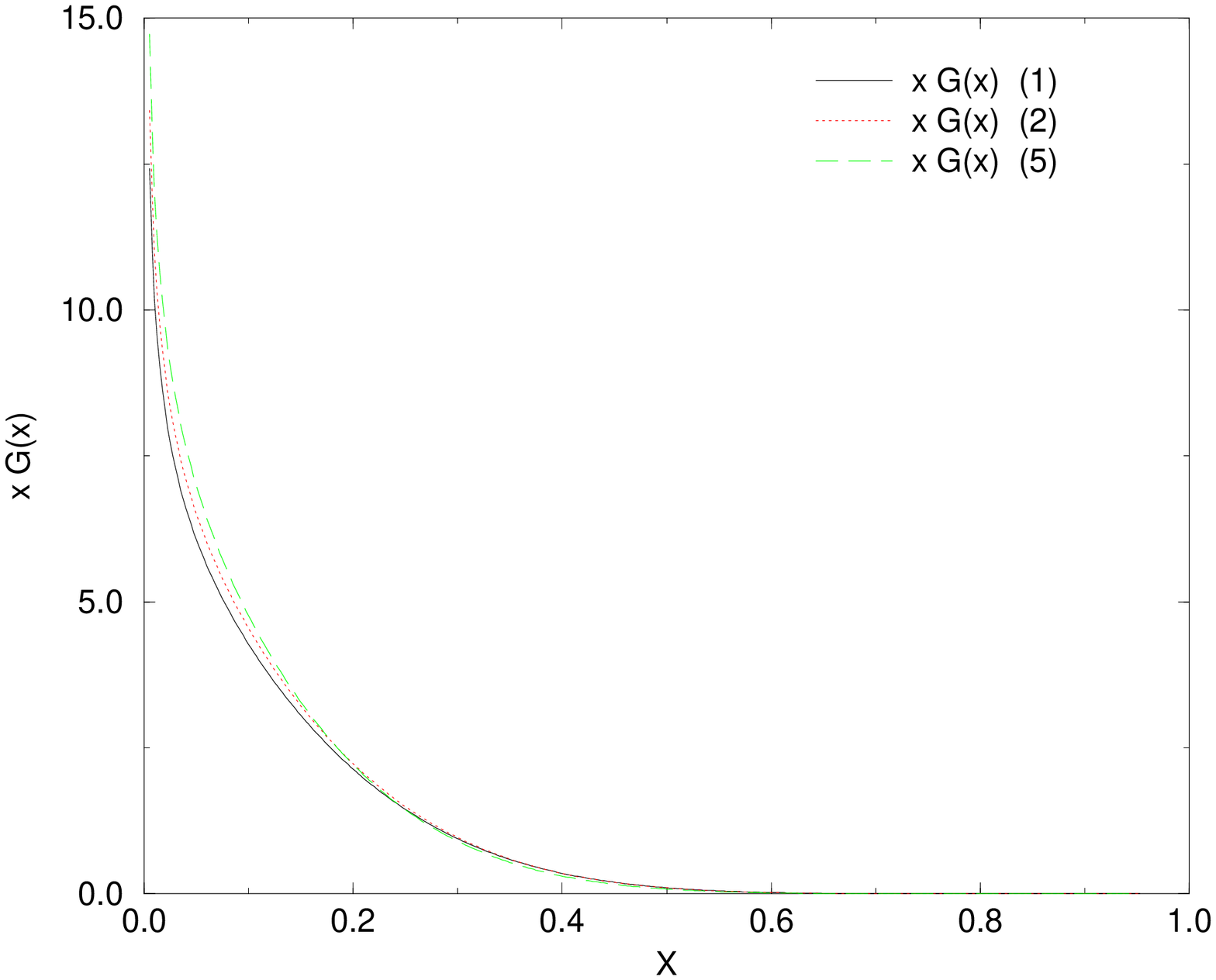}}
\caption{{ $x G(x)$ in the AP-ESAP evolution, 
with a squark mass $m_{\tilde{q}}=200$ GeV. The final evolution scales are 1,2, and 5 TeV respectively }}
\label{quark11}
\end{figure}

We also study the dependence of the evolution on the 
final evolution scale, keeping fixed the matching scale $m_{\tilde{q}}$. 
Figs.~\ref{quark10} and \ref{quark11} show the dependence of $x G(x)$ 
from the final evolution scale int the AP-ESAP evolution. 
We have chosen final evolution scales of 1,2, and 5 TeV respectively for both plots. 
The matching scales in the two plots are different. We have set $m_{\tilde{q}}= 20$ GeV and 200 GeV respectively. 
The plots show a similar pattern and are hardly distinguishable. 

\begin{figure}
\centerline{\includegraphics[angle=-90,width=.9\textwidth]{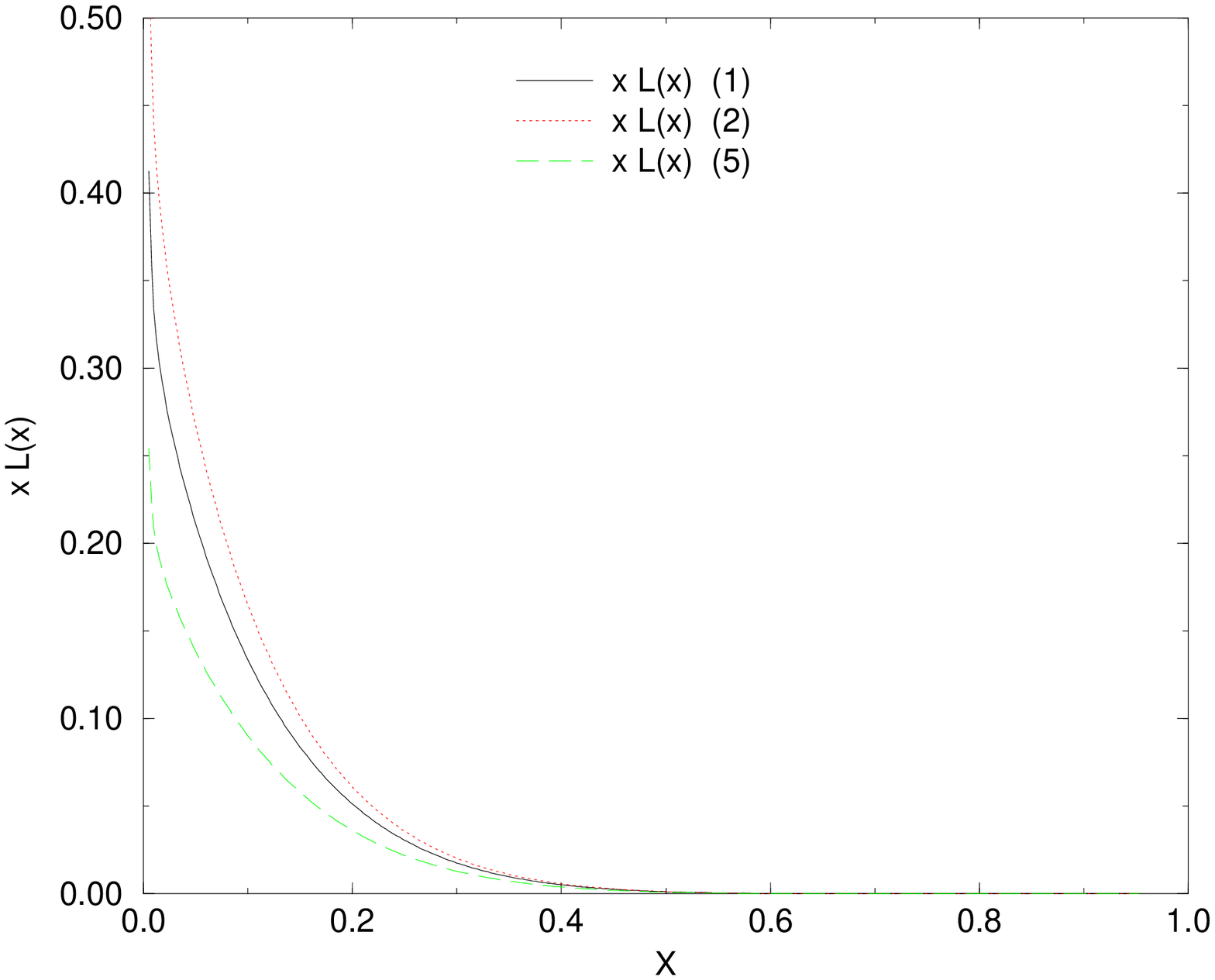}}
\caption{{ $x \lambda(x)$ in the AP-ESAP evolution, 
with a squark mass $m_{\tilde{q}}=200$ GeV. The final evolution scales are 1,2,and 5 TeV respectively }}
\label{quark12}
\centerline{\includegraphics[angle=-90,width=.9\textwidth]{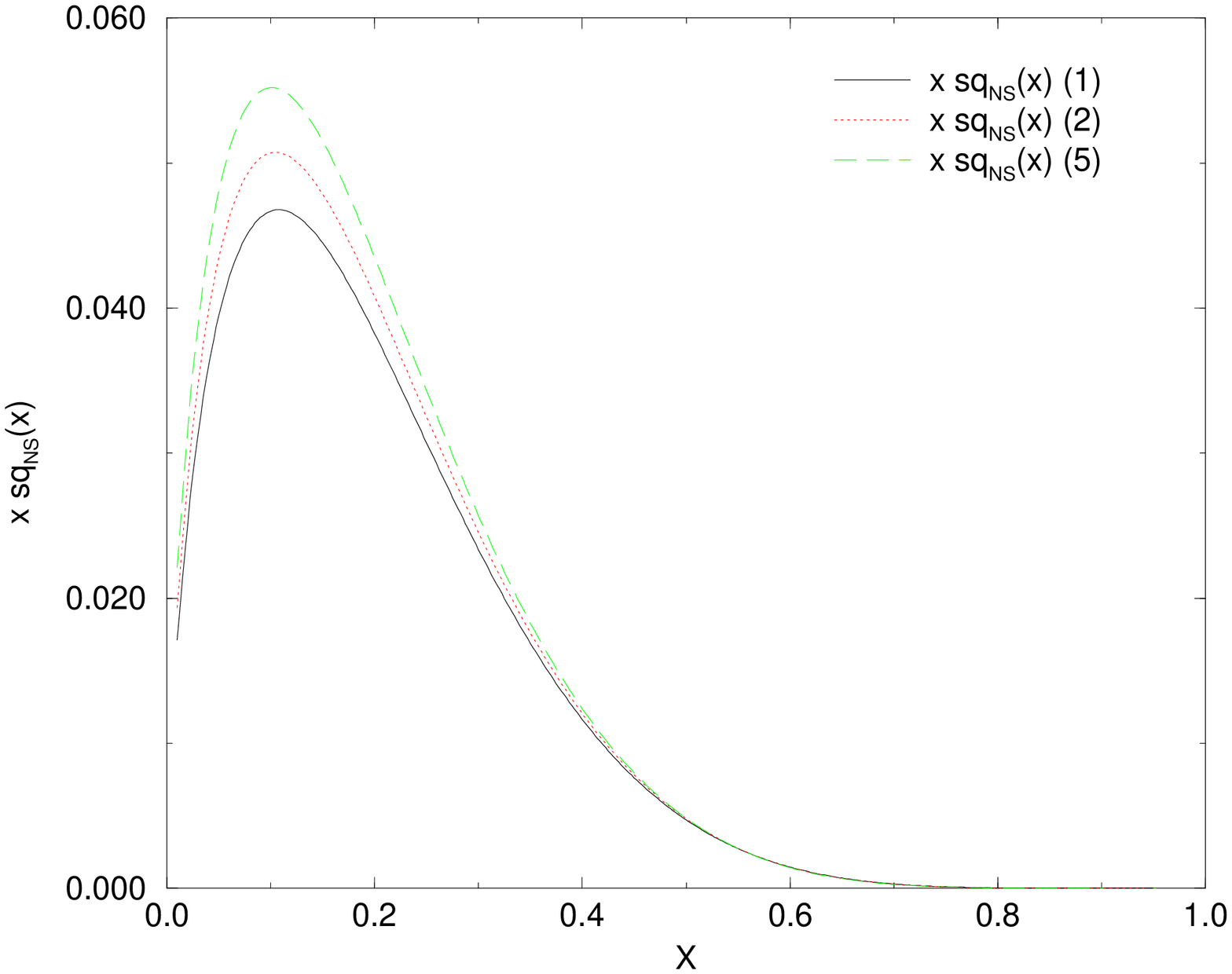}}
\caption{{ $x \tilde{q}^-(x)$ in the AP-ESAP evolution, 
with a squark mass $m_{\tilde{q}}=20$ GeV. The final evolution scales are 1,2, 
and 5 TeV respectively }}
\label{quark13}
\end{figure}

\begin{figure}
\centerline{\includegraphics[angle=-90,width=.9\textwidth]{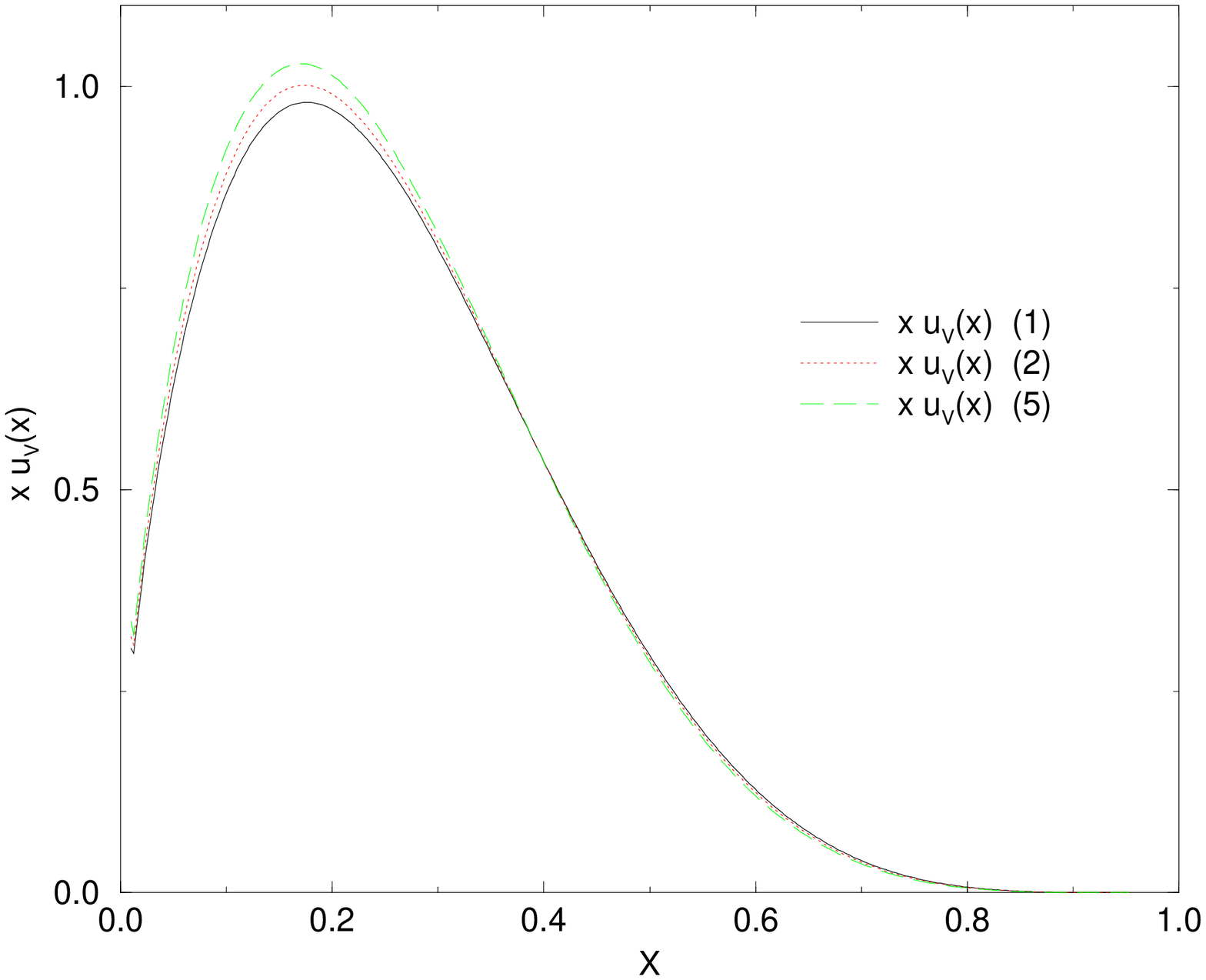}}
\caption{{ $x {q}^-(x)$ in the AP-ESAP evolution, 
with a squark mass $m_{\tilde{q}}=20$ GeV. The final evolution scales are 1,2, 
and 5 TeV respectively }}
\label{quark14}
\centerline{\includegraphics[angle=-90,width=.9\textwidth]{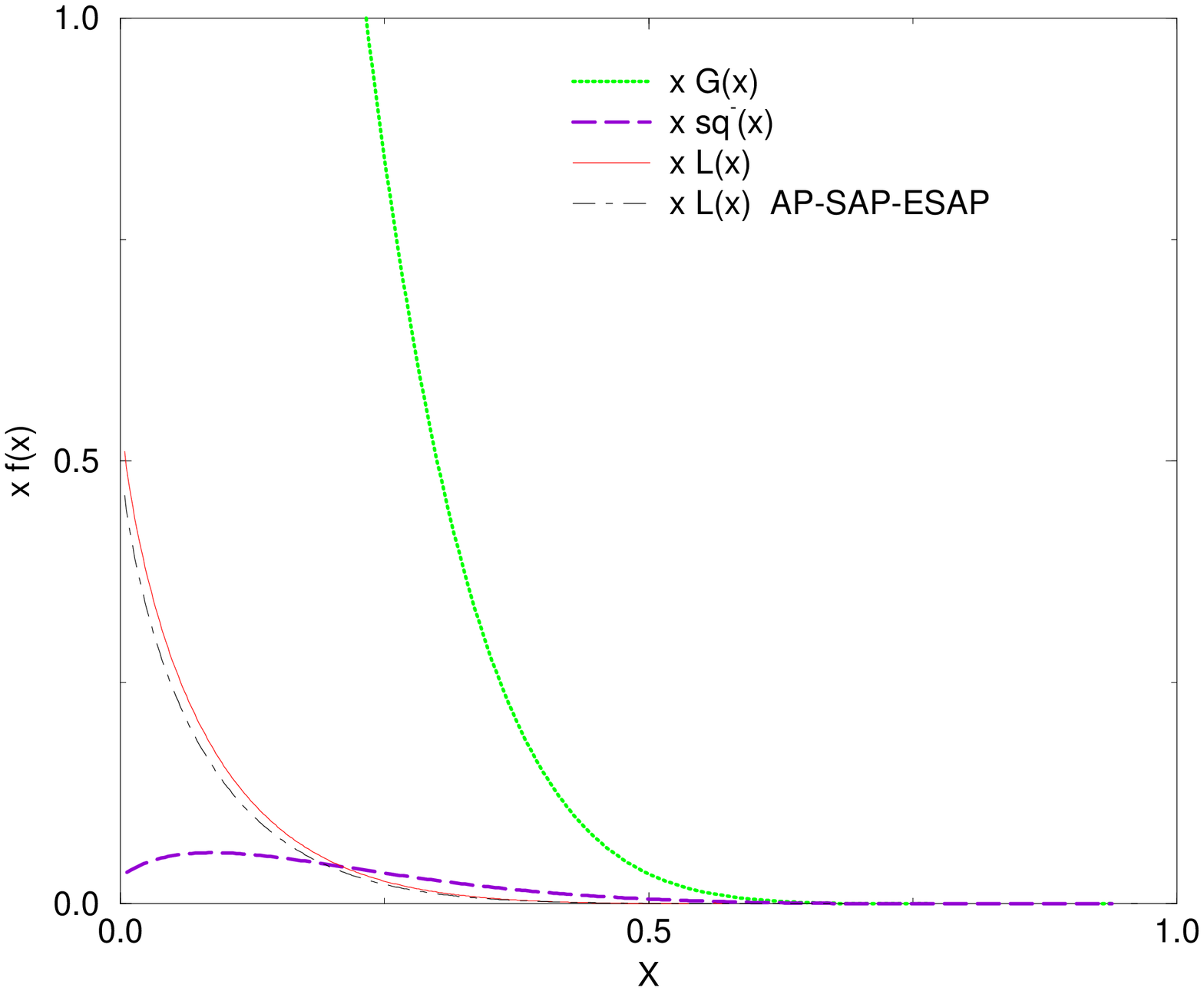}}
\caption{{ $x f(x)$ in the AP-ESAP evolution with a very large final scale 
$Q_f=10^3$ GeV and with 
a squark mass $m_{\tilde{q}}=100$ GeV. Shown are the non-singlet squark, 
the gluon and the gluino distributions for the AP-ESAP evolution. The gluino distribution for the AP-SAP-ESAP evolution is also shown 
(with $m_{2\lambda}=40$ GeV). }}
\label{quark15}
\end{figure}

In Fig.~\ref{quark12} we illustrate the dependence of 
the gluino density on the final evolution scales, chosen to be of 1, 2 and 5 TeV. 
As we increase 
the final scale $Q_f$ for a fixed $m_{\tilde{q}}$, 
the gluino distribution becomes steeper. Figs. \ref{quark13} and \ref{quark14} 
have been included to show the dependence of 
the the non-singlet distributions (valence) for squarks and quarks on the final evolution scale $Q_f$. The dependence is slightly more 
pronounced for squarks than for valence quarks. Finally, 
Fig.~\ref{quark15} shows the shapes of the gluino, 
gluon and non-singlet squark distributions when 
the final scale is very large. We have taken $Q_f=10^3$ GeV, 
corresponding to a very energetic collision. In this figure 
all the distributions -except one- have been obtained by an AP-ESAP evolution. One of the distributions is evolved using a general supersymmetric run (AP-SAP-ESAP) 
with an intermediate mass gluino (20 GeV) and a slightly heavier squark.  

\section{The General SUSY evolution}
The most general supersymmetric evolution is obtained by matching three 
regions: the AP region, the SAP region and the ESAP region. 
This is also a realistic approximation 
to the most general solution of the Renormalization Group Equation 
if gluinos  and squarks are widely separated in mass. 
In section 3 we have presented the (formal) solution to this most general evolution, where each stage of the evolution 
produces shapes of the distributions which are used as input for the next regions. 
To be specific, 
let's consider the most general $(Q_0,Q_1)_{AP}-(Q_1,Q_2)_{SAP}-(Q_2,Q_f)_{ESAP}$ supersymmetric 
evolution. We are allowed to vary the two matching scales $Q_1$ and $Q_2$. 
Partons become massless -this is the content of a step-approximation to the general solution- 
as soon as we step into a new region. In this respect, we should remark that threshold effects, 
in order to be kept fully into account, would require some knowledge on the way supersymmetry is broken (or restored) as we raise the energy scale. 
This goes beyond our actual understanding of the theory. 
From a perturbative viewpoint, additional anomalous dimensions 
-needed to perform a matching between the various regions - may be needed, 
especially in moving from the SAP region to the ESAP region. 
Similarly to QCD, it is expected that in leading order these effects are negligible.      
In the analysis that follows we will consider the following scenarios: a) a light gluino and a heavy squark; b) a heavier gluino and a much heavier squark 
c) a heavy gluino and a heavier squark at an extremely large evolution scale. Being the violations to scaling logarithmic, it is expected that the enhancements 
of the supersymmetric distributions will appear on an extremely large 
final evolution scale. As we are going to see, the enhancements -especially for gluinos- 
are larger at small-x and for smaller gluino masses. By varying 
the mass of the scalar quarks -here assumed to be degenerate, just for simplicity- 
we find that the modifications of the distributions are significant. 
We have kept $n_f=6$ in all the runs. 
Larger squark masses reduce drastically the maxima of the distributions, as illustrated below.

 In Fig.~\ref{xquark1} we show the dependence of the gluino density using a general 
AP-SAP-ESAP evolution. We have 2 matching parameters: $m_{2\lambda}$ and $m_{\tilde{q}}$. 
Here we have varied both matching scales and the changes induced on the 
distributions are shown to be small. A more pronounced modification 
on the amplitudes can be observed from Fig.~\ref{xquark2}, where we plot the non-singlet squark density for a light gluino and a heavier squark. The same pattern 
is observed in Fig.~\ref{xquark7} where we vary the mass of the squark at fixed 
gluino mass. The same effect of reduction of the distributions takes place 
as we raise the squark mass. In Fig.~\ref{xquark3} we compare 3 models of evolution: the 
general susy evolution, the partial susy evolution (AP-SAP) and the combined 
AP-ESAP evolution where we switch on an N=1 evolution 
right on top of the regular AP run. It is observed 
that a run with partial susy generates distributions of gluinos 
which are smaller than the corresponding fully supersymmetric ones. 
It is also evident that a mixed AP-SAP-ESAP evolution and an AP-ESAP run 
generates similar results given the parameters chosen in this case 
(light gluino, heavier squarks). In Fig.~\ref{xquark4} we plot the gluino distribution 
for a varying final evolution scale. The dependence is shown not to be significant. 
Fig.~\ref{xquark5} and \ref{xquark6} show a similar pattern 
when we vary both the gluino mass and the squark mass in the general susy evolution. 
In these two figures we have chosen to vary both matching scales 
(gluino and squark masses), without any appreciable modification on the result. 
We conclude that the gluino distribution is not much affected by changes 
in the paramters of the evolution. 
The same type of numerical study (for an AP-SAP-ESAP run) is carried out in Fig.~\ref{xquark8}, but for the singlet squark distribution. As we raise the squark mass, 
the distribution is lowered considerably. Here we have chosen a light 
gluino in the first range of the evolution. It is shown that the squark 
density is peaked at small-x and larger then a gluino density (light gluino) 
for the same x-value, at least as far as $m_{squark} < 100$ GeV.


\begin{figure}[thb]
\centerline{\includegraphics[angle=-90,width=.9\textwidth]{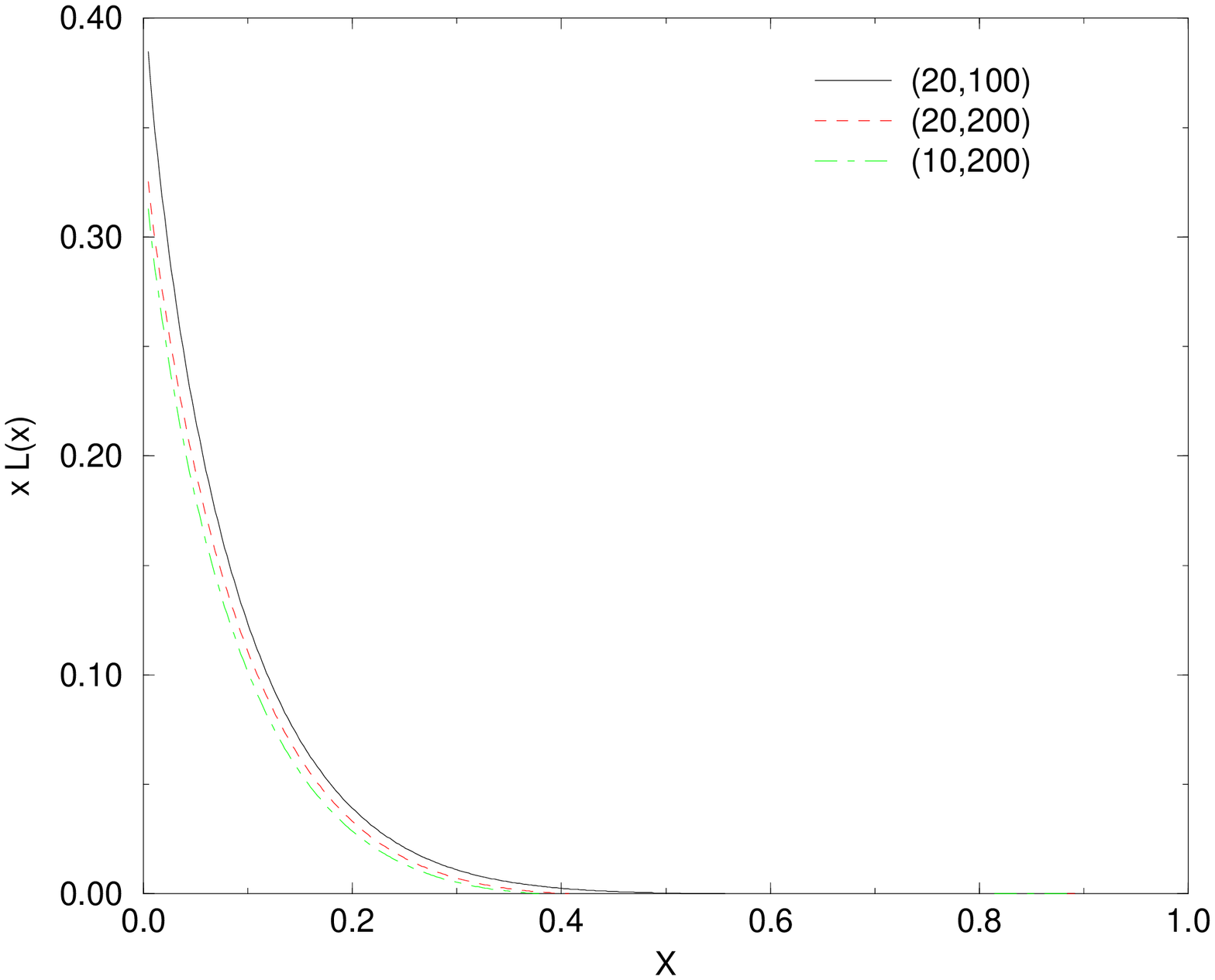}}
\caption{{ $x\lambda(x)$ using a general 
evolution. The matching parameters are denoted by two entries 
$(m_{2\lambda},m_{\tilde{q}})$ (in GeV) and $Q_f=500$ GeV}}
\label{xquark1}
\centerline{\includegraphics[angle=-90,width=.9\textwidth]{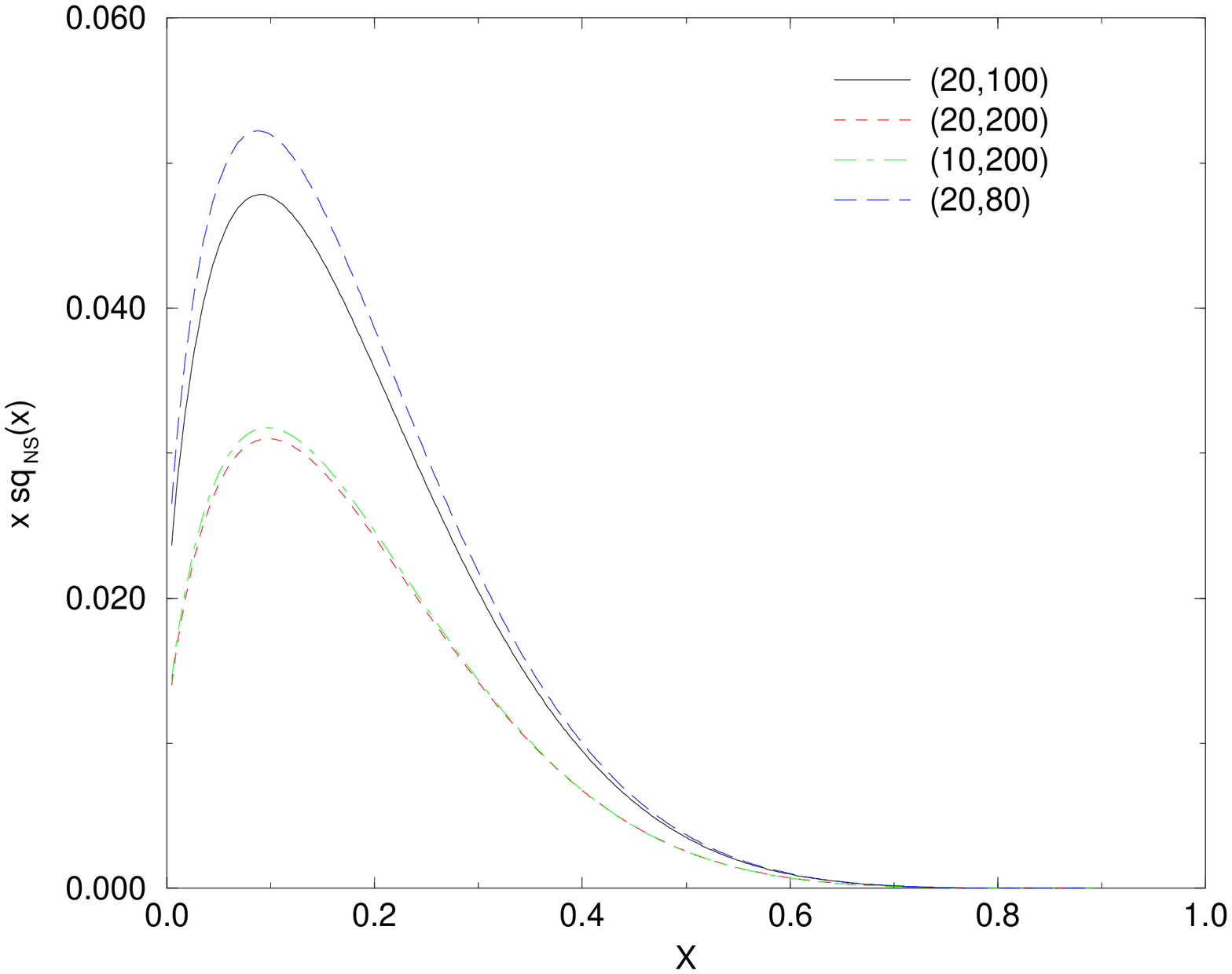}}
\caption{{ $x \tilde{q}^-(x)$ using a general 
evolution. The matching parameters are indicated as in Fig.~1 and $Q_f=500$ GeV}}
\label{xquark2}
\end{figure}

\begin{figure}[thb]
\centerline{\includegraphics[angle=-90,width=.9\textwidth]{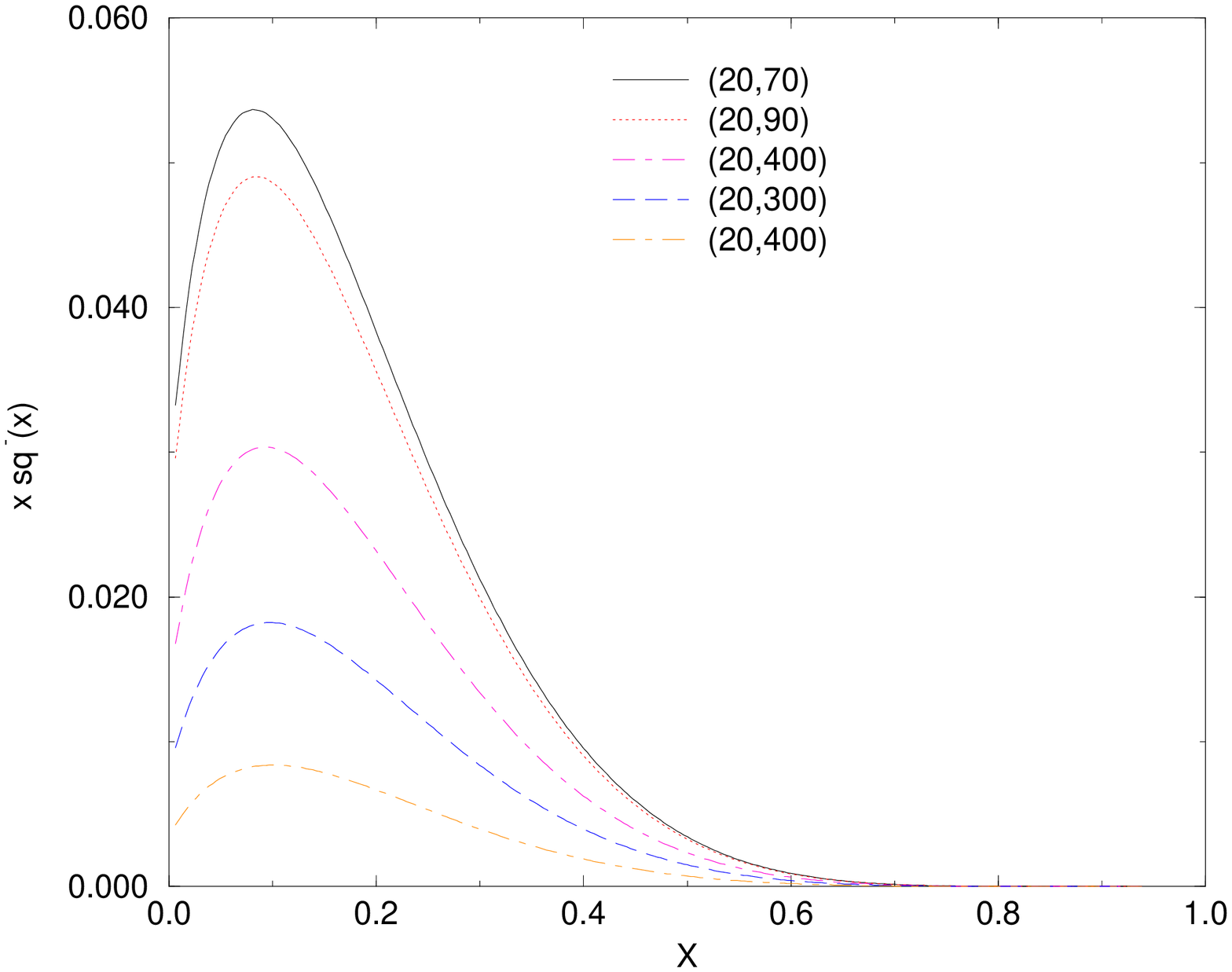}}
\caption{{Dependence of the singlet squark distribution on the squark mass
 for $m_{2\lambda}=20$ GeV and $m_{\tilde{q}}=70,90,200,300$ and 400 GeV. 
 We have chosen $Q_f=500$ TeV }}
\label{xquark7}
\centerline{\includegraphics[angle=-90,width=.9\textwidth]{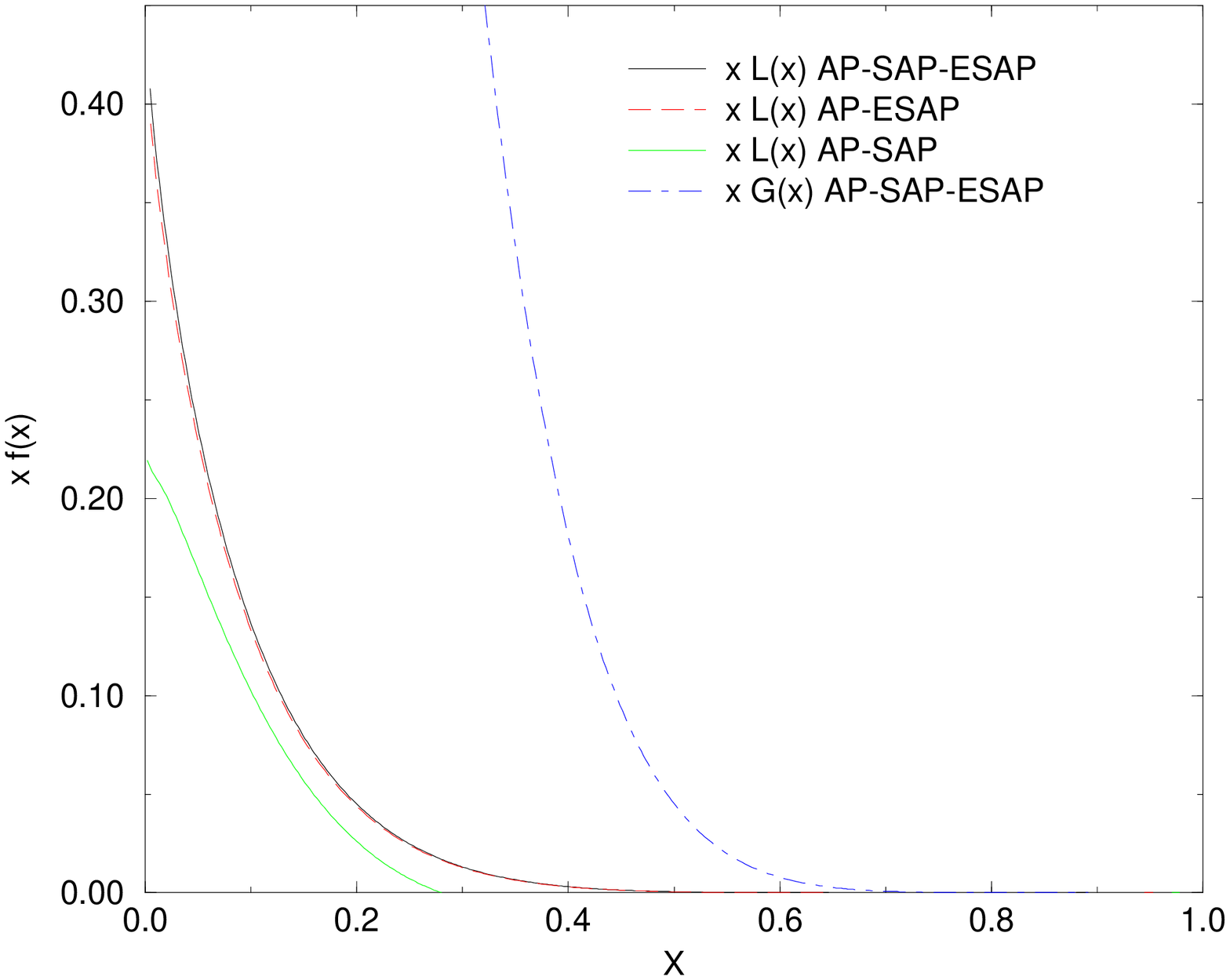}}
\caption{{ 3 gluino distributions according to the 3 possible evolution models,shown for $m_{2\lambda}=40$ GeV and $m_{\tilde{q}}$=100 GeV, 
and 1 gluon distribution. We have chosen $Q_f=500$ GeV. }}
\label{xquark3}
\end{figure}  

\begin{figure}[thb]
\centerline{\includegraphics[angle=-90,width=.9\textwidth]{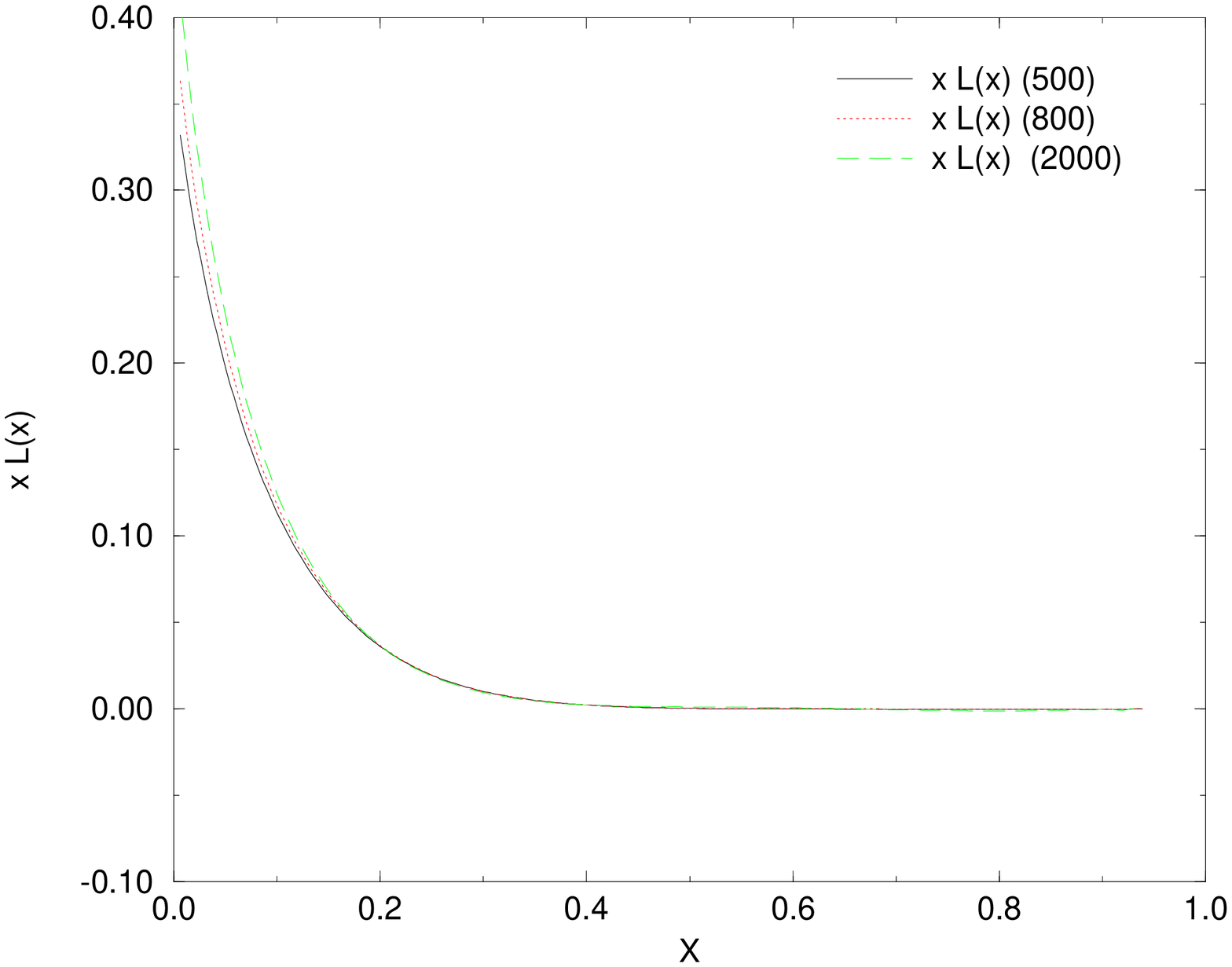}}
\caption{{ Gluino distribution $x \lambda(x)$ for $m_{2\lambda}=40$ GeV and 
$m_{\tilde{q}}=100$ GeV and $Q_f=500,800$ Gev and 2 TeV respectively  }}
\label{xquark4}
\centerline{\includegraphics[angle=-90,width=.9\textwidth]{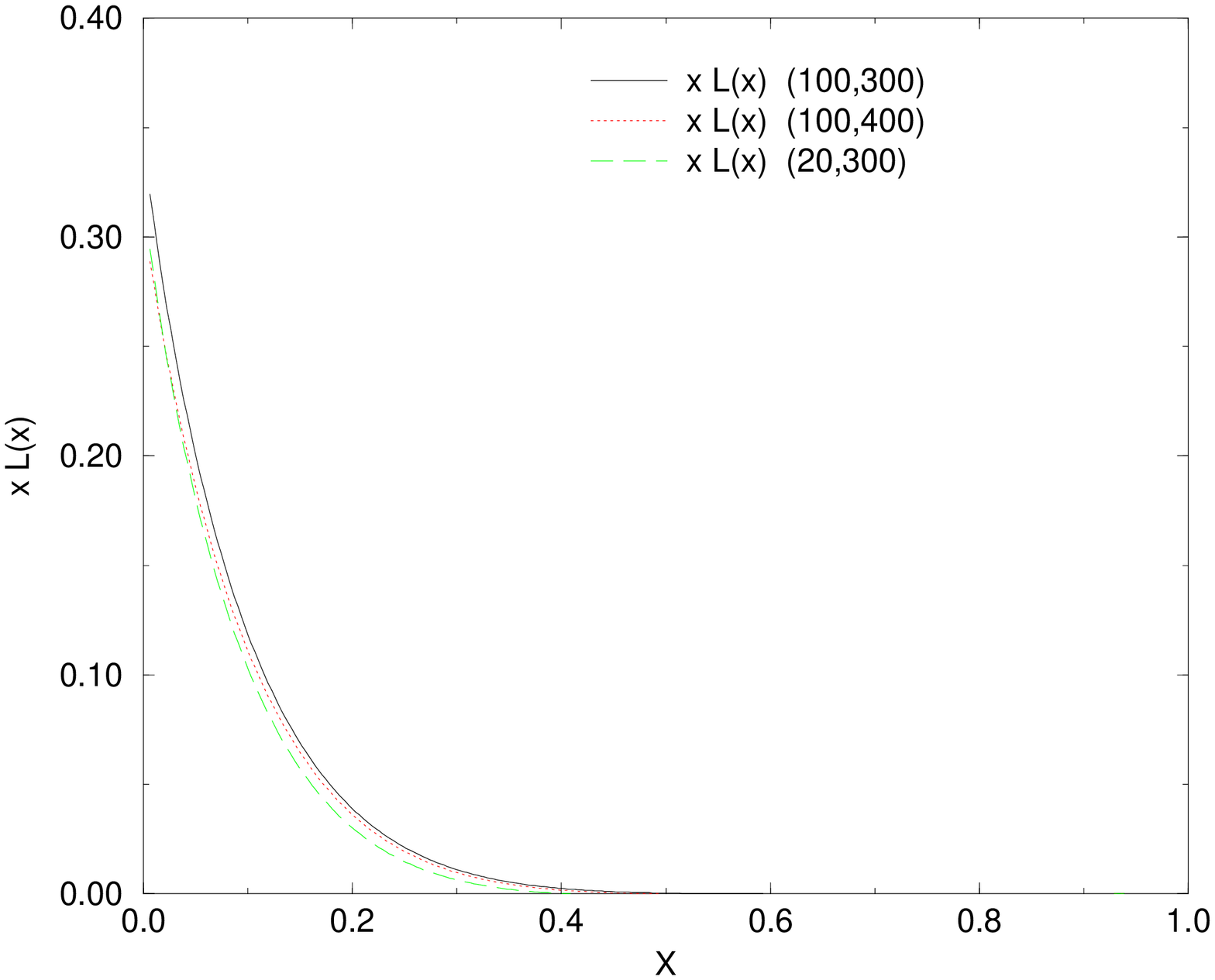}}
\caption{{ 3 gluino distributions according to the 3 possible evolution models,
 shown for $m_{2\lambda}=100$ and 20 GeV and for $m_{\tilde{q}}$=300,400 GeV. 
 We have chosen $Q_f=1$ TeV }}
\label{xquark5}
\end{figure}  

\begin{figure}[thb]
\centerline{\includegraphics[angle=-90,width=.9\textwidth]{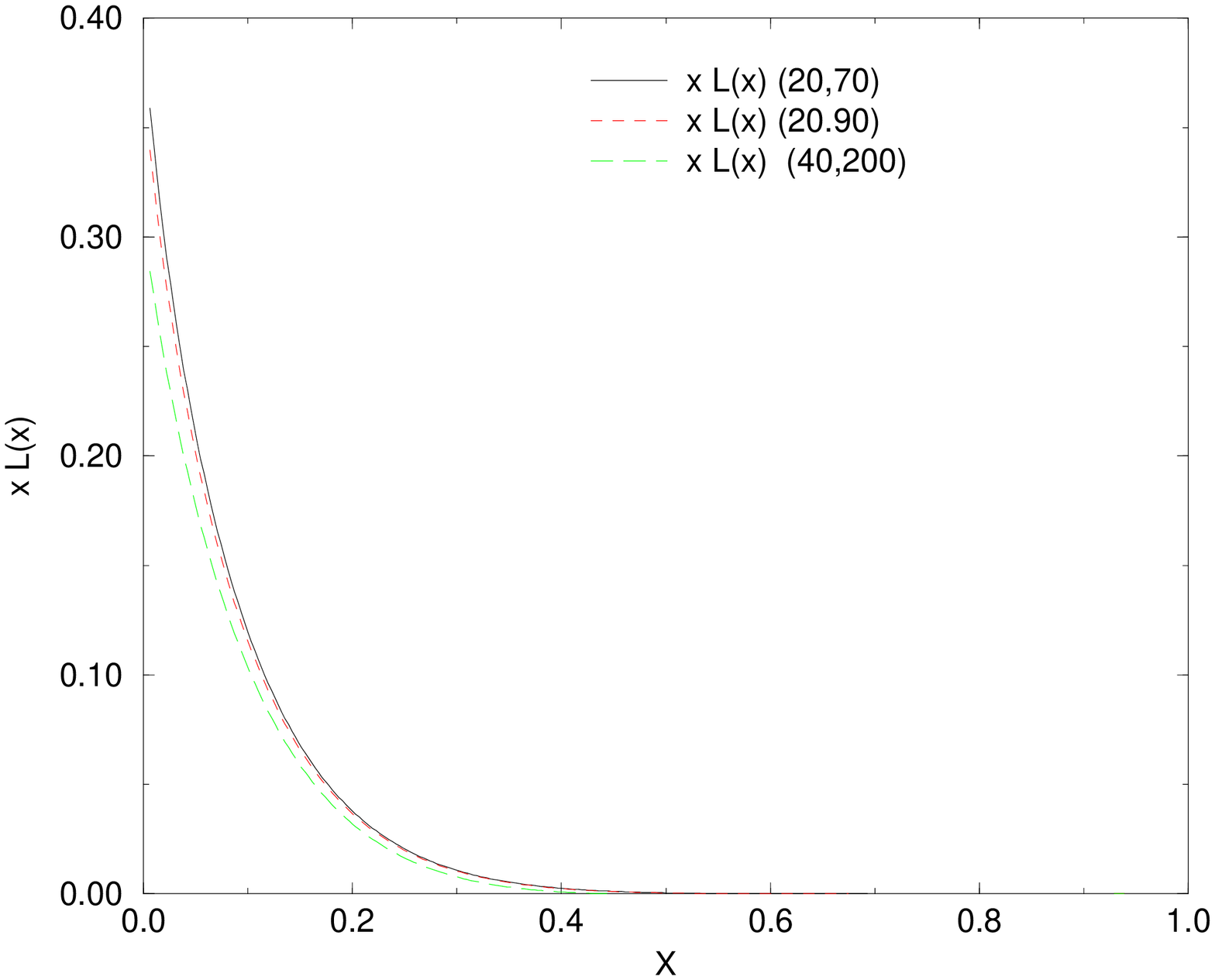}}
\caption{{ 3 gluino distributions according to the 3 possible evolution models,
 shown for $m_{2\lambda}=20$ GeV and $m_{\tilde{q}}=70,90$ and 200 GeV. 
 We have chosen $Q_f=500$ TeV }}
\label{xquark6}
\centerline{\includegraphics[angle=-90,width=.9\textwidth]{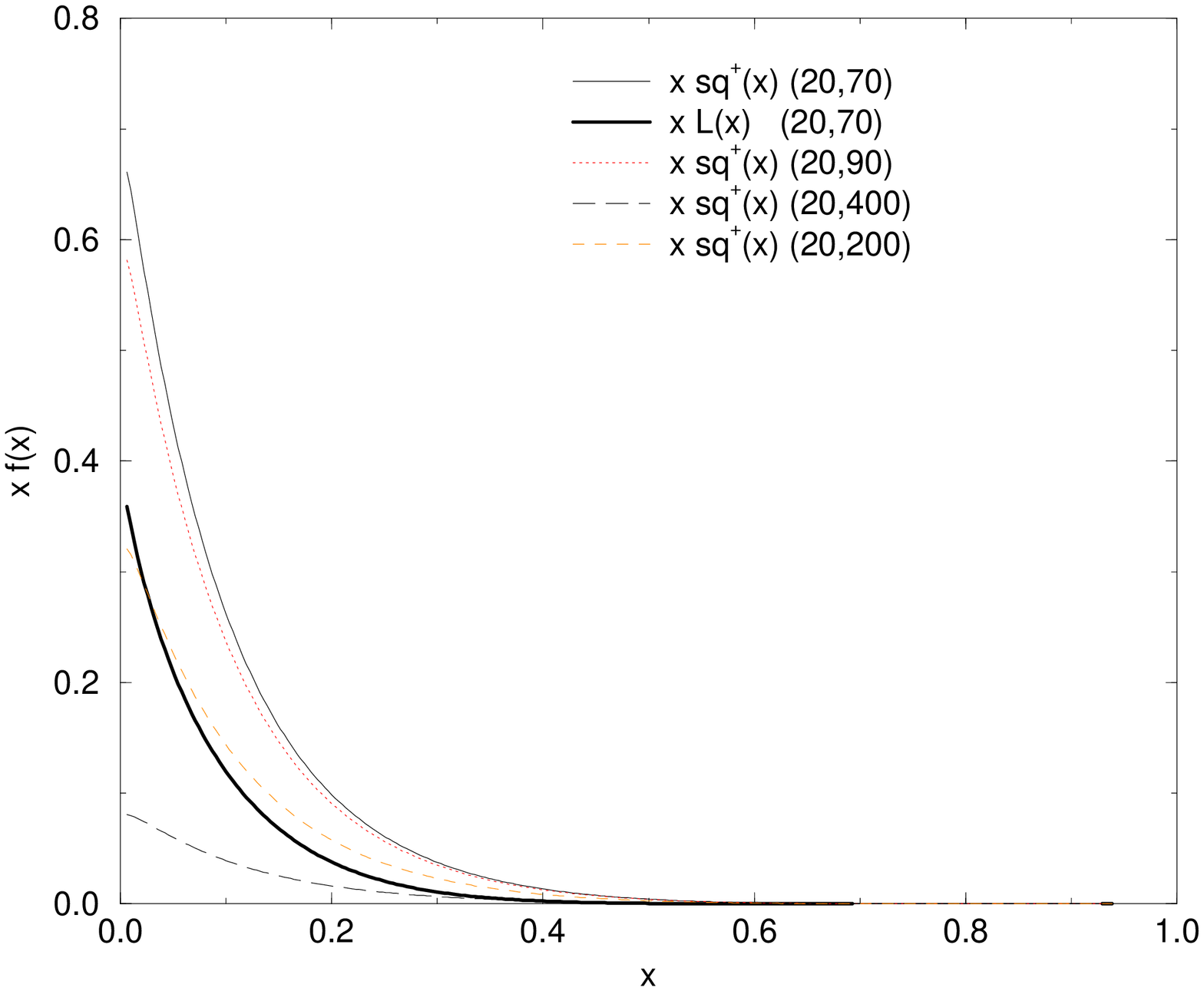}}
\caption{{ x $\tilde{q}^+(x)$ distributions for $m_{2\lambda}=40$ GeV and 
$m_{\tilde{q}}=70,90,200$ and 400 GeV respectively. Shown is also a gluino 
$x\lambda(x)$ distribution. We have chosen $Q_f=500$ TeV }}
\label{xquark8}
\end{figure}

\section{Conclusions}
Scaling violations induced by a supersymmetric evolution of the parton distributions have been studied in the context 
of a general supersymmetric evolution. The model distributions illustrated in this work have been 
generated within a radiative model, using the mixing of the 
QCD anomalous dimensions with the new states included 
in the $N=1$ as we cross each supersymmetric region in the evolution. Gluinos, and singlet squark distributions are strongly enhanced at small-x, 
have a faster decay at larger-x compared to the gluon distribution 
and show dependence on the parameters (the two matching scales) 
of the evolution. The dependence is less pronounced 
for gluinos compared to squarks. The regular 
QCD distributions also show a small dependence on the parameters 
of the evolution, although these effects might 
be of minor phenomenological relevence at current energies. 
We will presente elsewhere applications of 
these results to the computation of supersymmetric 
cross sections.

\section{Acknowledgements}
I thank the Theory Group at Oxford for financial support and hospitality 
and in particular Alon Faraggi and Subir Sarkar.

%%%%%%%%%%%%%%%%%%%%%%%%%%%%%

%%%%%%%%%%%%%%%%%%%%%%%%%%%%%%%%%%%%%%%%%%%%%%%%%%%%%%%
\section{Appendix}
The ESAP kernels are given by 
\beqa
P^{(0)}_{gg} & = & 2 C_A\left[ \frac{1}{(1-x)_+} + \frac{1}{x} -2 + 
x(1-x)\right] + \frac{\beta_0^{ES}}{2}\delta(1-x) \nonumber \\
P^{(0)}_{\lambda g}& = & n_\lambda\left[ 2 x - 1 \right]
\nonumber \\
P^{(0)}_{\lambda \lambda} & = & C_A \left[ \frac{2}{(1-x)_+} -1 -x +
 \right]+\left(\frac{3}{2} C_A -\frac{T_R}{2}\right)\delta(1-x) \nonumber \\
P^{(0)}_{\lambda q} & = & C_\lambda( 1-x) \nonumber \\
P^{(0)}_{\lambda s} & = &  C_F \nonumber \\
P^{(0)\,S}_{q g} &=& P^{(0)}_{q g}= n_f\left[ x^2 + (1-x)^2\right] \nonumber \\
P^{(0)}_{q \lambda} & = & n_f(1-x) \nonumber \\
P^{(0)\,S}_{qq} & = & C_F\left[{\left((1 + x^2)\over (1-x)\right)}
-\frac{1}{2}\delta(1-x) \right]_+\nonumber \\ 
& = & C_F\left( {2\over (1-x)_+} -1-x + \frac{3}{2}\delta(1-x)\right)\nonumber \\
P^{(0)}_{q s} & = & C_F \nonumber \\
P^{(0)}_{s g} & = & n_f \left[ 1 - \left( x^2 + (1-x)^2\right)\right] \nonumber \\
P^{(0)}_{s \lambda} & = & n_f x \nonumber \\
P^{(0)}_{s q} & = & C_F x \nonumber \\
P^{(0)}_{s s} & = &  C_F\left(\frac{2}{(1-x)_+} -2\right) + C_F\delta(1-x).\nonumber \\
\label{kernels}
\eeqa

The SAP kernels are given by 

\beqa
P^{(0)\,\,S}_{gg} & = & 2 C_A\left[ \frac{1}{(1-x)_+} + \frac{1}{x} -2 + 
x(1-x)\right] + \frac{\beta_0^S}{2}\delta(1-x) \nonumber \\
P^{(0) }_{g \lambda} & = & C_\lambda \left[ \frac{1 +(1-x)^2}{x}\right] \nonumber \\
P^{(0) S}_{g q} & = & P^{(0)}_{g q}=  C_F\left[ \frac{2}{x}  -2 + x  \right] \nonumber \\
P^{(0)}_{\lambda g}& = & n_\lambda\left[  1 - 2 x + 2 x^2 \right]\nonumber \\
P^{(0)}_{\lambda \lambda} & = & C_\lambda \left[ \frac{2}{(1-x)_+} -1 -x +
\frac{3}{2} \delta(1-x) \right]= C_\lambda\left(\frac{1 + x^2}{(1-x)}
\right)_+ \nonumber \\
P^{(0)}_{\lambda q} & = &0 \nonumber \\
P^{(0)\,S}_{q g} &=& P^{(0)}_{q g}= n_f\left[ x^2 + (1-x)^2\right] \nonumber \\
P^{(0)}_{q \lambda} & = & n_f(1-x) \nonumber \\
P^{(0)\,S}_{qq} & = & C_F\left[{(1 + x^2)\over (1-x)} \right]_+\nonumber \\ 
 & = & C_F\left( {2\over (1-x)_+} -1-x + \frac{3}{2}\delta(1-x)\right)\nonumber \\
\eeqa

\newpage


\begin{thebibliography}{99}
\bibitem{KR} C. Kounnas and D.A. Ross, Nucl. Phys. {\bf B214}:317, 1983.
\bibitem{Antoniadis} I. Antoniadis, C. Kounnas and R. Lacaze, Nucl. Phys. {\bf B211}:216, 1983.
\bibitem{CC1} C. Coriano', {\bf hep-ph/0009227}, submitted to Nucl. Phys. B. 
\bibitem{cteq} H.L. Lai et al, Phys. Rev. {\bf D55}:1280, 1997. 
Phys. Rev. {\bf D51}:4763, 1996.
\bibitem{BB} J. Botts and  J. Blumlein Phys.Lett.{\bf B325}:190-196,1994.
\bibitem{Contogouris} A.P. Contogouris and H. Tanaka, Phys. Rev. {\bf D31}
1638 (1985).  

\end{thebibliography}
\end{document}